\documentclass[12pt]{iopart}
\usepackage{times,amsmath}
\usepackage{bm}
\usepackage{epsfig}
\usepackage{amsfonts}
\usepackage{subfigure}
\usepackage{color}

\begin{document}

\title{On the Minimal Scattering Response of PEC Cylinders in a Dielectric Cloak}

\author{Constantinos A. Valagiannopoulos, Pekka Alitalo and Sergei A. Tretyakov\\
\address{Department of Radio Science and Engineering,\\
         School of Electrical Engineering, Aalto University, Finland,\\
         P.O. Box 13000, FI-00076 Aalto}
\ead{\{konstantinos.valagiannopoulos, pekka.alitalo, sergei.tretyakov\}@aalto.fi}}

\begin{abstract}
Recently, it was shown that an infinite perfectly conducting (PEC)
cylinder can be nearly perfectly cloaked from normally incident
electromagnetic waves using a single-layer homogeneous dielectric
cladding. Here we study the electromagnetic response of such
structures with the goal to understand the main mechanisms
underpinning this cloaking phenomenon. We introduce a simple model of
the cloaked PEC cylinder, replacing it by an omnidirectional
electric-line scatterer and a bipolar magnetic one; accordingly, the
far field is determined in a compact closed form. The analysis of
the results shows that the optimal cloaking regime corresponds to
the frequency point where the total electric moment is drastically
mitigated and thus the radiation pattern of the device resembles
that of a magnetic-dipole line. In the vicinity of the optimal
cloaking frequency we observe the response close to that of Huygens'
pairs of electric and magnetic line scatterers.
\end{abstract}

\maketitle

\section{Introduction}
Cloaking devices render objects partially or wholly invisible to
parts of the electromagnetic spectrum by drastically reducing the
total scattering cross section of the objects. Due to the
obvious interest in potential applications and due to their
intriguing nature, cloaks have attracted considerable attention from
both practical experts and theoretical scientists during the
previous decade. In the pioneering works \cite{Leonhardt_Science, Pendry_Science}, the
control of electromagnetic fields in order to not to interact with
an object is theoretically formulated. They use optical conformal mapping techniques 
and complex coordinate
transformations which require exotic inhomogeneous and anisotropic materials. On the
other hand, actual structures realizing broadband electromagnetic
cloaking with, e.g., transmission lines, are presented in the
papers \cite{Pekka_PIEEE, New_TAP}, where experimental validation is
provided. Furthermore, the so-called ``scattering cancellation''
technique has been introduced in \cite{alu_ScatteringCancellation},
where drastic reduction of the scattering response of a specific
dielectric object is accomplished with a dielectric covering layer.

In this paper we  consider cloaking of infinite perfectly-conducting
(PEC) cylinders using the simplest possible configuration: a uniform
dielectric cover \cite{Ours_PRB}. A PEC circular cylinder is covered
by a single cladding and is excited by a plane wave (whose electric field is parallel 
to the cylinder axis), formulating a
two-dimensional (2D) boundary-value problem. This geometric
configuration is the same as that for scattering cancellation
cloaking of dielectric cylinders \cite{alu_ScatteringCancellation},
but the physical phenomena leading the the cloaking effect are
different and the known design methods for scattering-cancellation
cloaks cannot be used for this polarization \cite{alu_Retrospection}. In fact, the
equations for the required thickness and material parameters of the
cloaking cladding have no solution for the case of a PEC cylinder in
the center \cite{alu_ScatteringCancellation,alu_Retrospection}.
However, similar devices have been analyzed in electrostatic
\cite{Zharov_Electrostatics}, in quasi-static
\cite{Milton_Resonance} and in electrodynamic \cite{Ours_PRB}
regimes and satisfying cloaking performances have been achieved. It
is therefore important to understand the phenomenon and develop a
predictive analytical model, enabling the design of cloaks for elongated
conductive objects.

Here we introduce an approximate model that can be used in
explaining and interpreting the aforementioned interesting behavior.
In particular, we note that the scattering pattern of cloaked PEC
cylinders in the vicinity of the optimal cloaking frequency is
similar to simple dipolar objects. In addition, we expect that the
optimal cloaking regime should correspond either to minimization of
induced lowest-order moment(s)  or to the electric and magnetic
moments forming a  Huygens' pair with the minimum scattering in the
forward direction.  This brings us to the idea to model the
cylindrical volume by one electric and one magnetic dipole line
located along the central axis. The corresponding moments are
evaluated analytically from the canonical field solution and the
scattered field of the equivalent structure is expressed in a
compact form. After validating the model by comparing the data
originating from the exact solution with that given by the
approximate model, we discuss the frequency response of the
equivalent system. Certain conclusions related to the principle of
the device operation and justification of the cloaking phenomenon
are drawn. Some preliminary results of this study have been reported
in a conference  \cite{Ours_ICEAA}.

\section{Electromagnetic Fields}

\begin{figure}[ht]
\centerline{\includegraphics[scale=0.8]{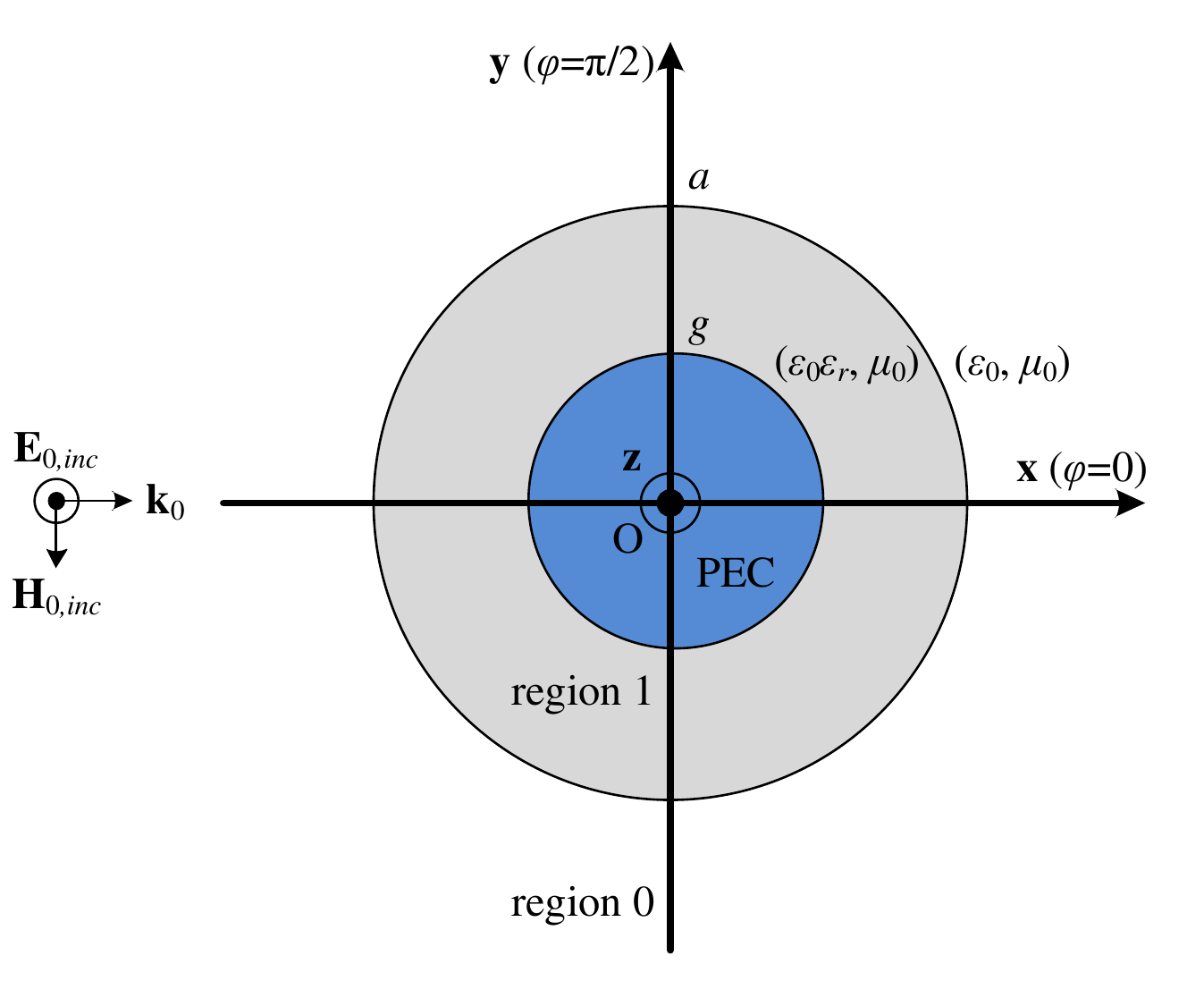}}
\caption{A simple 2D structure of a PEC cylinder of radius $g$ covered by a magnetically inert layer of thickness $(a-g)$ and relative permittivity $\epsilon_r$, is illuminated by a plane wave.}
\label{fig:Fig1}
\end{figure}

Let us consider a simple two-dimensional (2D) structure depicted in
Fig.~\ref{fig:Fig1}. An infinite PEC cylinder of the radius $g$ is
covered by a single isotropic and uniform dielectric layer
of the thickness $(a-g)$. The cladding layer (Region~1) is filled
with a magnetically inert material of the relative permittivity
$\epsilon_{r}$. The whole structure is placed in vacuum (Region~0, the
permittivity $\epsilon_0$, and the permeability $\mu_0$). The axis
of the cylinder coincides with the $\textbf{z}$ axis of the
cylindrical coordinate system $(\rho, \phi, z)$, which can be used
interchangeably with the corresponding Cartesian one $(x, y, z)$.
This cylindrical configuration is illuminated by a
$\textbf{z}$-polarized (TE or TM$_z$ in an alternative definition)
plane wave of unitary electric-field magnitude propagating along the
positive $\textbf{x}$ semi-axis \cite{PlaneWaveExpansion}:
\begin{eqnarray}
\textbf{E}_{0,inc}=\textbf{z}E_{z0,inc}(\rho,\phi)=\textbf{z}\sum_{n=0}^{+\infty}A_0(n)J_n(k_0\rho)\cos(n\phi),
\label{eq:IncidentElectricField}
\end{eqnarray}
where $A_0(n)=\frac{2}{1+\delta_{n0}}j^{-n}$, $k_0=2\pi f \sqrt{\epsilon_0 \mu_0}$ is the free-space wavenumber and $f$ is the operational frequency. The notation $J_n$ concerns the Bessel function of $n$-th order, while the symbol $\delta_{n0}$ corresponds to the Kronecker delta. The harmonic time dependence is of the form $e^{+j2\pi ft}$ and suppressed throughout the analysis.

Due to the 2D nature of the studied configuration and the isotropy
of the materials, the (secondary) electric field in each of the two
regions (0, 1) is $\textbf{z}$-polarized and can be expressed in the
following forms:
\begin{eqnarray}
E_{z0,scat}(\rho, \phi)=\sum_{n=0}^{+\infty}B_0(n)H_n^{(2)}(k_0\rho)\cos(n\phi),
\label{eq:ElectricField0}
\end{eqnarray}
\begin{eqnarray}
E_{z1}(\rho, \phi)=\sum_{n=0}^{+\infty}\left[A_1(n)J_n(k \rho)+B_1(n)H_n^{(2)}(k \rho)\right]\cos(n\phi),
\label{eq:ElectricField1}
\end{eqnarray}
where $H_n^{(2)}=J_n-jY_n$ is the Hankel function of $n$-th order.
The wavenumber in region~0 is defined as $k=k_0\sqrt{\epsilon_{r}}$.

By imposing the necessary boundary conditions around the cylindrical
surfaces $\rho=g,a$, the  unknown coefficient sequences $B_0(n),
A_1(n), B_1(n)$ of (\ref{eq:ElectricField0}),
(\ref{eq:ElectricField1}) are readily determined. The azimuthal
magnetic component in region~1 is expressed as:
\begin{eqnarray}
H_{\phi 1}(\rho,\phi)=-\frac{j}{k_0\zeta_0}k \sum_{n=0}^{+\infty}\left[A_1(n)J'_n(k \rho)+B_1(n)H_n^{(2)\prime}(k \rho)\right]\cos(n\phi),
\label{eq:MagneticField1}
\end{eqnarray}
where $\zeta_0=\sqrt{\mu_0/\epsilon_0}=120\pi~\Omega$ is the wave
impedance of vacuum. Based on these field quantities, the equivalent
dipole moments can be rigorously evaluated. Furthermore, the
far-field response of the cylindrical structure is determined with
the use of Hankel's expansion for large arguments:
$H_n^{(2)}(k_0\rho)\sim \sqrt{\frac{2}{\pi k_0
\rho}}e^{-j\left(k_0\rho-\frac{n\pi}{2}-\frac{\pi}{4}\right)},
k_0\rho\rightarrow+\infty$. In particular,
\begin{eqnarray}
E_{z0,scat}(\rho, \phi)\sim \sqrt{\frac{2}{\pi k_0\rho}}e^{-j(k_0\rho-\frac{\pi}{4})}\sum_{n=0}^{+\infty}B_0(n)j^n\cos(n\phi), \quad k_0\rho\rightarrow+\infty.
\label{eq:ScatteringFarField}
\end{eqnarray}

\section{Electromagnetic Moments}
\subsection{Equivalent Current}
The surface current induced at the PEC surface $\rho=g$ is given by:
\begin{eqnarray}
\textbf{K}(\phi)=\textbf{z}H_{\phi 1}(g, \phi).
\label{eq:SurfaceCurrent}
\end{eqnarray}
By definition, the polarization current over the cross sections of
the cladding layer is written as follows:
\begin{eqnarray}
\textbf{J}_{pol}(\rho, \phi)=\textbf{z} j\frac{k_0}{\zeta_0}(\epsilon_{r}-1)E_{z1}(\rho, \phi).
\label{eq:PolarizationCurrent}
\end{eqnarray}
Consequently, the total induced current which is the source of the
scattering field by the cylindrical structure possesses the form:
\begin{eqnarray}
\textbf{J}(\rho, \phi)=\textbf{K}(\phi)\delta(\rho-g)+\textbf{J}_{pol}(\rho, \phi)\nonumber\\
=\textbf{z}H_{\phi 1}(g, \phi)\delta(\rho-g)+\textbf{z} j\frac{k_0}{\zeta_0}(\epsilon_{r}-1)E_{z1}(\rho, \phi),
\label{eq:EquivalentCurrent}
\end{eqnarray}
where $\delta$ is the Dirac function.

\subsection{Electric Moment}
The electric moment $\textbf{p}$ of electric current density
\textbf{J} existing in a volume $(V)$ is defined in
\cite{ElectricMomentP} and can be written in the 2D case as an
integral over the cross-section surface $(S)$:
\begin{eqnarray}
\textbf{p}=\frac{1}{j2\pi f}\iiint_{(V)}\textbf{J}~dV \Rightarrow \textbf{p}^*=\frac{1}{j2\pi f}\iint_{(S)}\textbf{J}~dS,
\label{eq:ElectricMoment}
\end{eqnarray}
where $\textbf{p}^*$ is the electric moment per unit length (p.u.l.)
of the $\textbf{z}$ axis. If one substitutes expression
(\ref{eq:EquivalentCurrent}) into formula (\ref{eq:ElectricMoment}),
only the 0-th term from the harmonic series
(\ref{eq:ElectricField1}), (\ref{eq:MagneticField1}) survives;
accordingly, the following expression is obtained:
\begin{eqnarray}
\textbf{p}^*=\textbf{z}\frac{2\pi}{k_0^2\zeta_0 c}
\left\{\begin{array}{c}k_0^2(\epsilon_{r}-1)\left[A_1(0)V_J(g, a, k)+B_1(0)V_H(g, a, k)\right]\\
                      -k g\left[A_1(0)J'_0(k g)+B_1(0)H_0^{(2)\prime}(k g)\right]
\end{array}\right\},
\label{eq:ElectricMomentExplicit}
\end{eqnarray}
where $c=\frac{1}{\sqrt{\epsilon_0\mu_0}}$ is the speed of light.
The integrals of the Bessel and Hankel functions are analytically
evaluated:
\begin{eqnarray}
V_J(\chi, \psi, k)=\int_{\chi}^{\psi}J_0(k\rho)\rho d\rho=\frac{\psi J_1(k \psi)-\chi J_1(k \chi)}{k},
\label{eq:VJFunction}
\end{eqnarray}
\begin{eqnarray}
V_H(\chi, \psi, k)=\int_{\chi}^{\psi}H_0^{(2)}(k\rho)\rho d\rho=\frac{\psi H_1^{(2)}(k \psi)-\chi H_1^{(2)}(k \chi)}{k}.
\label{eq:VHFunction}
\end{eqnarray}
This quantity $\textbf{p}^*=\textbf{z}~p_z^*$ (in Coulomb) expresses
mainly how powerful is the total induced electric current for given
device dimensions and the frequency. It can be considered as a 2D
electric dipole line (or an electric current line) along the central
axis of the structure, which models the electric response of the
whole cylindrical configuration.

\subsection{Magnetic Moment}
The magnetic moment $\textbf{m}$ of a volumetric distribution of
electric current density \textbf{J} existing in volume $(V)$ is
defined in \cite{MagneticMomentM} and can be rewritten in the 2D
case as a surface integral over the cross section $(S)$:
\begin{eqnarray}
\textbf{m}=\frac{1}{2}\iiint_{(V)}\textbf{r}\times \textbf{J}~dV \Rightarrow \textbf{m}^*=\frac{1}{2}\iint_{(S)}\textbf{r}\times \textbf{J}~dS,
\label{eq:MagneticMoment}
\end{eqnarray}
where $\textbf{r}={\bm \rho}\rho+\textbf{z}z$ is the observation
vector in cylindrical coordinates and $\textbf{m}^*$ is  the
magnetic moment per unit length (p.u.l.) of the $\textbf{z}$ axis.
Obviously, the vector $(\textbf{r}\times \textbf{J})$ has
exclusively an azimuthal (${\bm \phi}$) component. Due to the orthogonality of the
cylindrical functions, only the 1-st term from the harmonic series
(\ref{eq:ElectricField1}), (\ref{eq:MagneticField1}) survives.
Writing the unit vector along ${\bm \phi}$ in the Cartesian
coordinates: ${\bm \phi}=-\textbf{x}\sin\phi+\textbf{y}\cos\phi$, we
see that the vector of the magnetic dipole moment is parallel to the
$\textbf{y}$ axis (the term $\textbf{x}\sin\phi$ gives an odd
function with zero contribution to the integral). Accordingly, one
obtains the result:
\begin{eqnarray}
\textbf{m}^*=-\textbf{y}\frac{j\pi}{2k_0\zeta_0}
\left\{\begin{array}{c}k_0^2(\epsilon_{r}-1)\left[A_1(1)W_J(g, a, k)+B_1(1)W_H(g, a, k)\right]\\
                      -k g^2 \left[A_1(1)J'_1(k g)+B_1(1)H_1^{(2)\prime}(k g)\right]
\end{array}\right\}.
\label{eq:MagneticMomentExplicit}
\end{eqnarray}
Again, the related integrals of the Bessel and Hankel functions are
rigorously determined as follows:
\begin{eqnarray}
W_J(\chi, \psi, k)=\int_{\chi}^{\psi}J_1(k\rho)\rho^2 d\rho=\frac{\psi^2 J_2(k \psi)-\chi^2 J_2(k \chi)}{k},
\label{eq:WJFunction}
\end{eqnarray}
\begin{eqnarray}
W_H(\chi, \psi, k)=\int_{\chi}^{\psi}H_1^{(2)}(k\rho)\rho^2 d\rho =W_J(\chi, \psi, k) \nonumber \\
-\frac{j}{2}\left\{\begin{array}{c}
 \psi^3 M\left(\left[\begin{array}{c}-\frac{1}{2} \\ -1\end{array}\right], \left[\begin{array}{cc} -\frac{1}{2} & \frac{1}{2} \\ -\frac{3}{2} & -1 \end{array}\right], \frac{k\psi}{2}, \frac{1}{2}\right)\\
-\chi^3 M\left(\left[\begin{array}{c}-\frac{1}{2} \\ -1\end{array}\right], \left[\begin{array}{cc} -\frac{1}{2} & \frac{1}{2} \\ -\frac{3}{2} & -1 \end{array}\right], \frac{k\chi}{2}, \frac{1}{2}\right)\end{array}\right\},
\label{eq:WHFunction}
\end{eqnarray}
where $M$ is the Meijer function \cite{MeijerGFunction}.  This
quantity $\textbf{m}^*=\textbf{y}~m_y^*$ (in Ampere$\cdot$meter)
expresses mainly how powerful is the spatial variation of the
induced electric current over the cross section of the device for
given device dimensions and the frequency. It is a 2D magnetic
dipole line positioned along the central axis of the structure.

In the following we will see that this simple model adequately
represents the response of the cloaked cylinder near the optimal
operational frequency despite the fact that the variations of the
fields inside the cloak are quite significant (its electrical size
is considerable due to high permittivity of the cladding cylinder).

\subsection{Radiation Fields}
The radiated electric field  $\textbf{E}_p$ by an electric moment
$\textbf{p}^*$ p.u.l. of the axis $\textbf{z}$ in a 2D configuration
is written as follows \cite{Orfanidis_RadiationFields}:
\begin{eqnarray}
\textbf{E}_p(\rho, \phi)=\frac{1}{\epsilon_0}\nabla\times \left[\nabla G(\rho, \phi)\times \textbf{p}^* \right],
\label{eq:PElectricField}
\end{eqnarray}
where $G$ is the scalar Green's function in two dimensions,
expressed in cylindrical coordinates. It corresponds to the field at
a point $(\rho,\phi)$ developed from a singular source at $(P,
\Phi)$. In our case, the position of this dipole coincides with the
$\textbf{z}$ axis $(P=0)$ of the coordinate system; therefore, the
respective expression takes the form \cite{My_Green}:
\begin{eqnarray}
G(\rho, \phi)=-\frac{j}{4}H_0^{(2)}(k_0\rho).
\label{eq:GreenFunction}
\end{eqnarray}
Similarly, the radiated electric field $\textbf{E}_m$ produced by a
magnetic-moment line $\textbf{m}^*$ p.u.l. at the axis $\textbf{z}$
in a 2D configuration is given by \cite{Orfanidis_RadiationFields}:
\begin{eqnarray}
\textbf{E}_m(\rho, \phi)=-j2\pi f\mu_0\nabla G(\rho, \phi)\times \textbf{m}^*.
\label{eq:MElectricField}
\end{eqnarray}

If one takes into account the polarization of the moments
$(\textbf{p}^*=\textbf{z}~p_z^*, \textbf{m}^*=\textbf{y}~m_y^*)$,
substitutes Green's function (\ref{eq:GreenFunction}) and adds the
two electric fields (\ref{eq:PElectricField}),
(\ref{eq:MElectricField}), the total developed field
$(\textbf{E}_p+\textbf{E}_m)$ from the moments couple possesses a
single $\textbf{z}$ component given by:
\begin{eqnarray}
E'_{z0,scat}(\rho, \phi)=\frac{k_0^2\zeta_0}{4}\left[m_y^* H_1^{(2)}(k_0\rho)\cos\phi-j c p_z^* H_0^{(2)}(k_0\rho)\right].
\label{eq:RadiatedField}
\end{eqnarray}
The corresponding far-field expression reads:
\begin{eqnarray}
E'_{z0,scat}(\rho, \phi)\sim\sqrt{\frac{2}{\pi k_0\rho}}e^{-j(k_0\rho-\frac{\pi}{4})}\frac{k_0^2\zeta_0}{4j}
\left[c p_z^* - m_y^* \cos\phi\right], \quad k_0\rho\rightarrow+\infty.
\label{eq:RadiatedFarField}
\end{eqnarray}
The two electric fields $E_{z0,scat}(\rho, \phi)$ and 
$E'_{z0,scat}(\rho, \phi)$ represent the same physical quantity, namely
the field scattered by the cylindrical structure under the plane
wave excitation. The difference is that the first one
(the unprimed, denoted by: $E_{z0,scat}(\rho, \phi)$) is evaluated from the explicit forms of
the fields determined through the rigorous canonical solution of the
scattering problem, while the second one (the primed, denoted by: $E'_{z0,scat}(\rho, \phi)$)
is computed from the approximate model of two scattering dipole
lines.

\section{Total Scattering Quantities}

By repeating the same procedures described above, one can readily find the corresponding field quantities $\tilde{E}_{z0,scat}(\rho, \phi)$, $\tilde{E}'_{z0,scat}(\rho, \phi)$ which describe the situation of a bare PEC cylinder of radius $g$ in the absence of the cladding (our solution for $\epsilon_r=1$). In this way, one can define the normalized total scattering widths in each case as follows:
\begin{eqnarray}
\sigma_{norm}=\lim_{k_0\rho\rightarrow+\infty}\frac{\int_0^{2\pi} |E_{z0,scat}(\rho, \phi)|^2 d\phi}{\int_0^{2\pi} |\tilde{E}_{z0,scat}(\rho, \phi)|^2 d\phi},\\
\label{eq:NormalizedTotalScatteringWidth}
\sigma'_{norm}=\lim_{k_0\rho\rightarrow+\infty}\frac{\int_0^{2\pi} |E'_{z0,scat}(\rho, \phi)|^2 d\phi}{\int_0^{2\pi} |\tilde{E}'_{z0,scat}(\rho, \phi)|^2 d\phi}.
\label{eq:NormalizedTotalScatteringWidthPrimed}
\end{eqnarray}
These quantities measure the reduction of total scattering from the
PEC cylinder due to cloaking cladding  (when $\sigma_{norm},
\sigma'_{norm}\rightarrow 0$, we have perfect cloaking effect). Also
here  $\sigma_{norm}$ and  $\sigma'_{norm}$ give the normalized
scattering widths computed from the rigorous solution and the dipole
moments, respectively.

The forward scattering theorem \cite{ForwardScatteringTheorem} is a well-known lemma, frequently used in electromagnetics, which relates the scattered field by a lossless object $E_{z0,scat}(\rho, \phi), E'_{z0,scat}(\rho, \phi)$, along the forward direction (namely the one of the incident plane wave, here: $\phi=0$) with the total scattered power from the object. In particular, this power (expressed in Watt) is given by:
\begin{eqnarray}
P_{scat}=-\frac{\sqrt{2}}{k_0\zeta_0}\Re\left[\sum_{n=0}^{+\infty}B_0(n)j^n\right],\\
\label{eq:ScatteredPower}
P'_{scat}=-\frac{k_0}{2\sqrt{2}}\Im\left[c p_z^*-m_y^*\right],
\label{eq:RadiatedPower}
\end{eqnarray}
respectively to which method we adopt (the full-wave solution or the moments model). In other words, in the case of a perfect cloak, the electric quantity $\Im[c p_z^*]$ should be equal and opposite to the magnetic quantity $\Im[-m_y^*]$. The two sources $\Im[c p_z^*], \Im[-m_y^*]$ are cooperating with each other to produce mutually neutralized scattering.

\section{Numerical Results}
\subsection{Parameters}
Before proceeding with the discussion  of the results, we need to
define the value ranges of the problem parameters. As the reference
operational frequency of the cloak, we assume $f_0=3\cdot 10^8$~Hz
corresponding to unitary free-space wavelength $\lambda_0=1$~m. The
physical size of the considered structure is kept constant
throughout the numerical simulations to simplify the procedure and
highlight the effect of variations of the other parameters. 
More specifically, we take $g=0.05\lambda_0$ and $a=0.08\lambda_0$. This
assumption  does not affect the generality of our investigation
because similar dependencies are observed for other dimensions.
Since the operational frequency does not differ significantly from
$f_0$, the electrical dimension of the PEC cylinder is close to 10\%
of the free-space operational wavelength. In other words, cloaking
(even not perfectly) this conducting volume is not a trivial task
since its electrical size is far from infinitesimal even in terms of
the free-space wavelength.

\subsection{Graphs}

\begin{figure}[ht]
\centering
\subfigure[]{\includegraphics[scale =0.5]{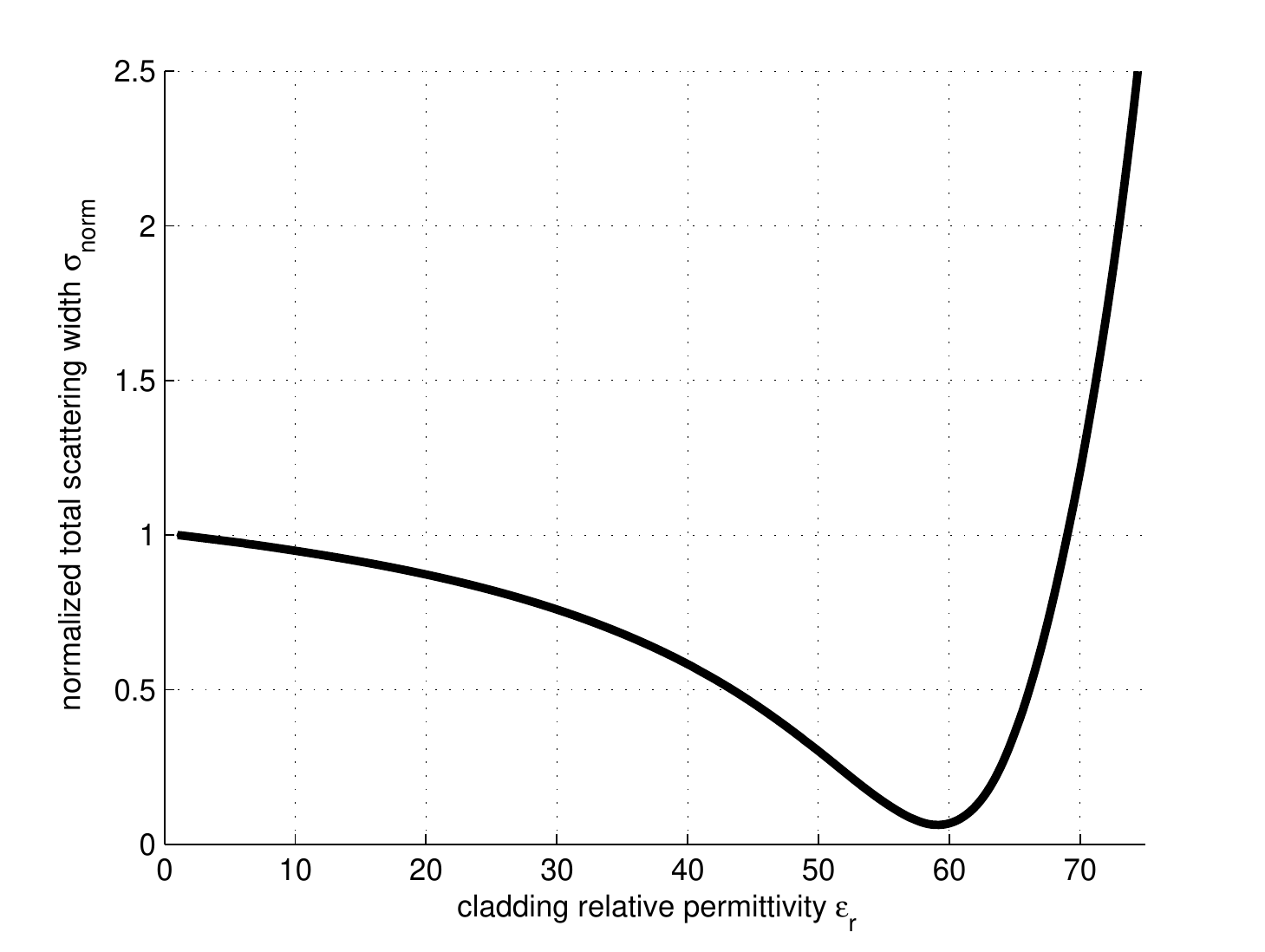}
   \label{fig:Fig2a}}
\subfigure[]{\includegraphics[scale =0.5]{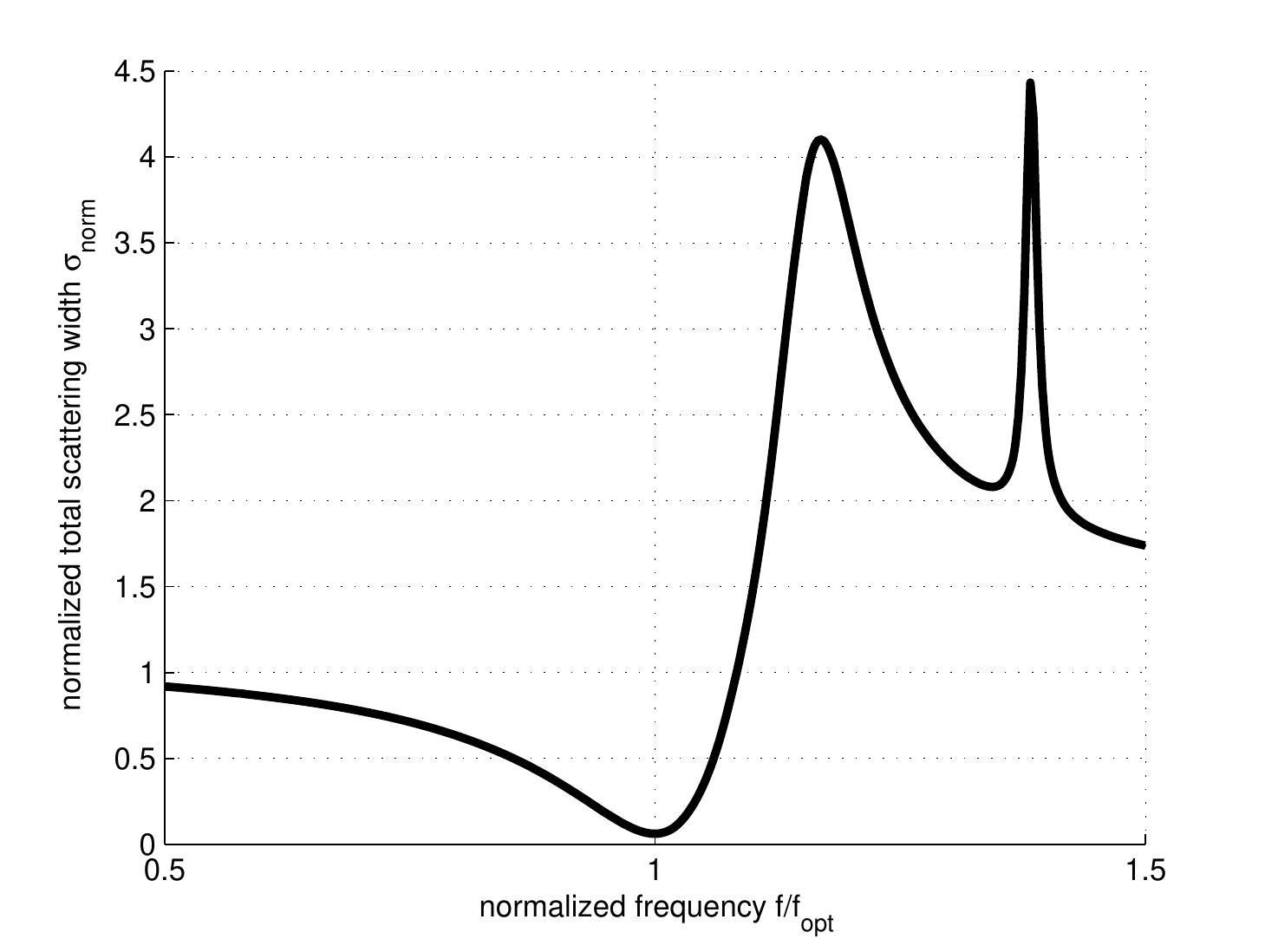}
   \label{fig:Fig2b}}
\caption{The normalized total scattering width $\sigma_{norm}$ as function of: (a) the cladding relative permittivity $\epsilon_r$ ($f=f_0$) and (b) the normalized operating frequency $f/f_{opt}$ ($\epsilon_r=60$). Plot parameters: $f_0=c/\lambda_0=3\cdot 10^8$ Hz, $f_{opt}=0.992f_0$, $g/\lambda_0=0.05$, $a/\lambda_0=0.08$.}
\label{fig:Figs2}
\end{figure}

In Fig.~\ref{fig:Fig2a} we sweep the relative permittivity of the
cladding $\epsilon_r$ and evaluate the normalized total scattering
width using the exact field formulas. The operating frequency is
kept constant $f=f_0$ and the optimal cloaking result (the minimum
$\sigma_{norm}$) is obtained for $\epsilon_r\cong 60$. Therefore,
the dielectric constant is kept fixed to this value hereinafter to
emphasize the frequency response of the device. Needless to say
that additional minima are recorded for higher frequencies, but we
are interested in the cloaking regime at the lowest possible
frequency (not so high permittivity and electrical
size). In Fig.~\ref{fig:Fig2b} we plot the same quantity
$\sigma_{norm}$ with respect to the frequency. The optimal frequency
$f_{opt}$ is very close to $f_0$ ($f_{opt}=0.992f_0$) but not
exactly equal to  that since the choice $\epsilon_{r}=60$ does not
correspond to the optimal case of Fig.~\ref{fig:Fig2a}.
Additionally, the shape of the two curves in Figs.~\ref{fig:Figs2}
are similar since the horizontal axis corresponds to a common
quantity: the effective electrical distance inside the cladding.

\begin{figure}[ht]
\centering
\subfigure[]{\includegraphics[scale =0.5]{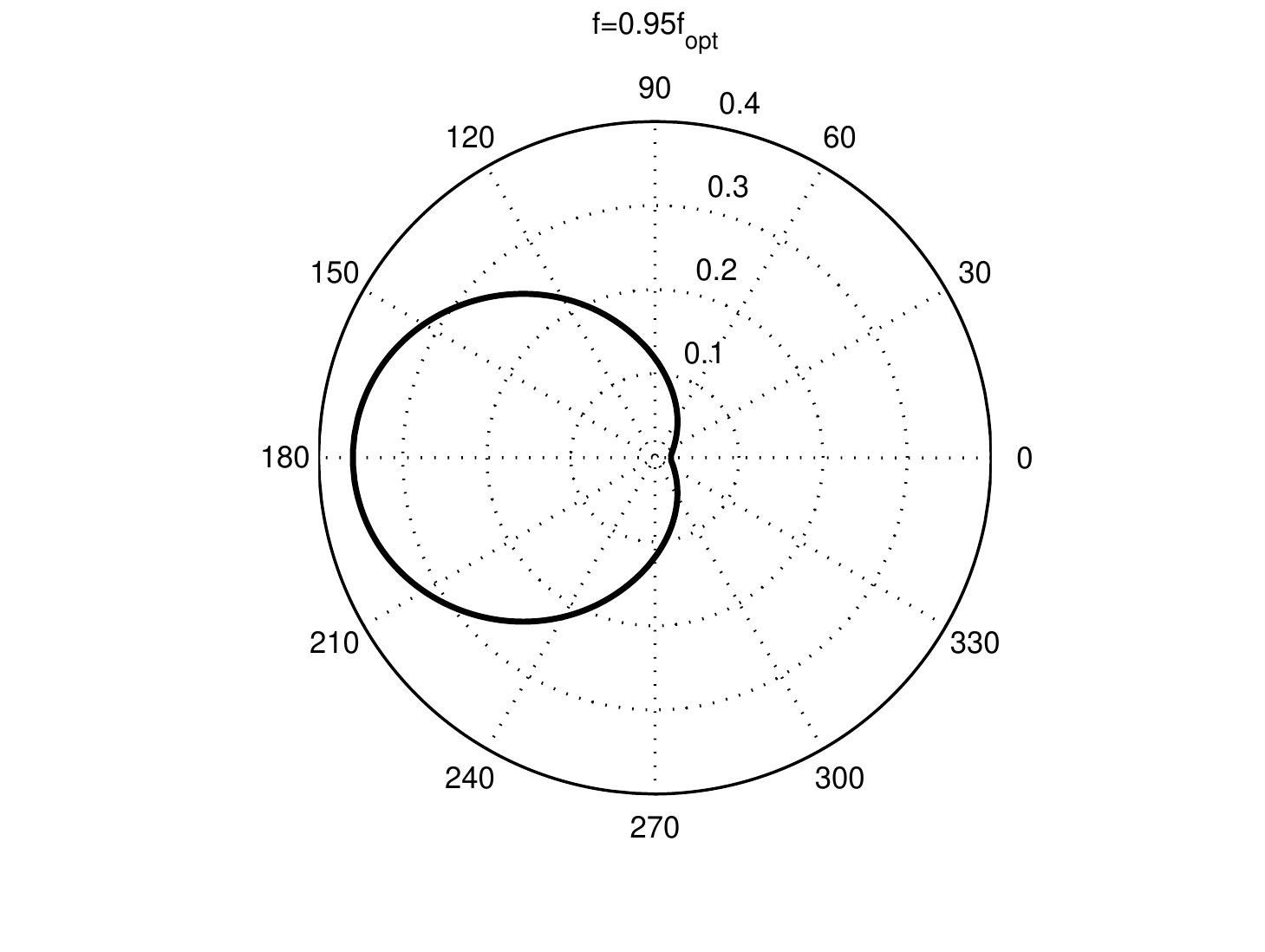}
   \label{fig:Fig3a}}
\subfigure[]{\includegraphics[scale =0.5]{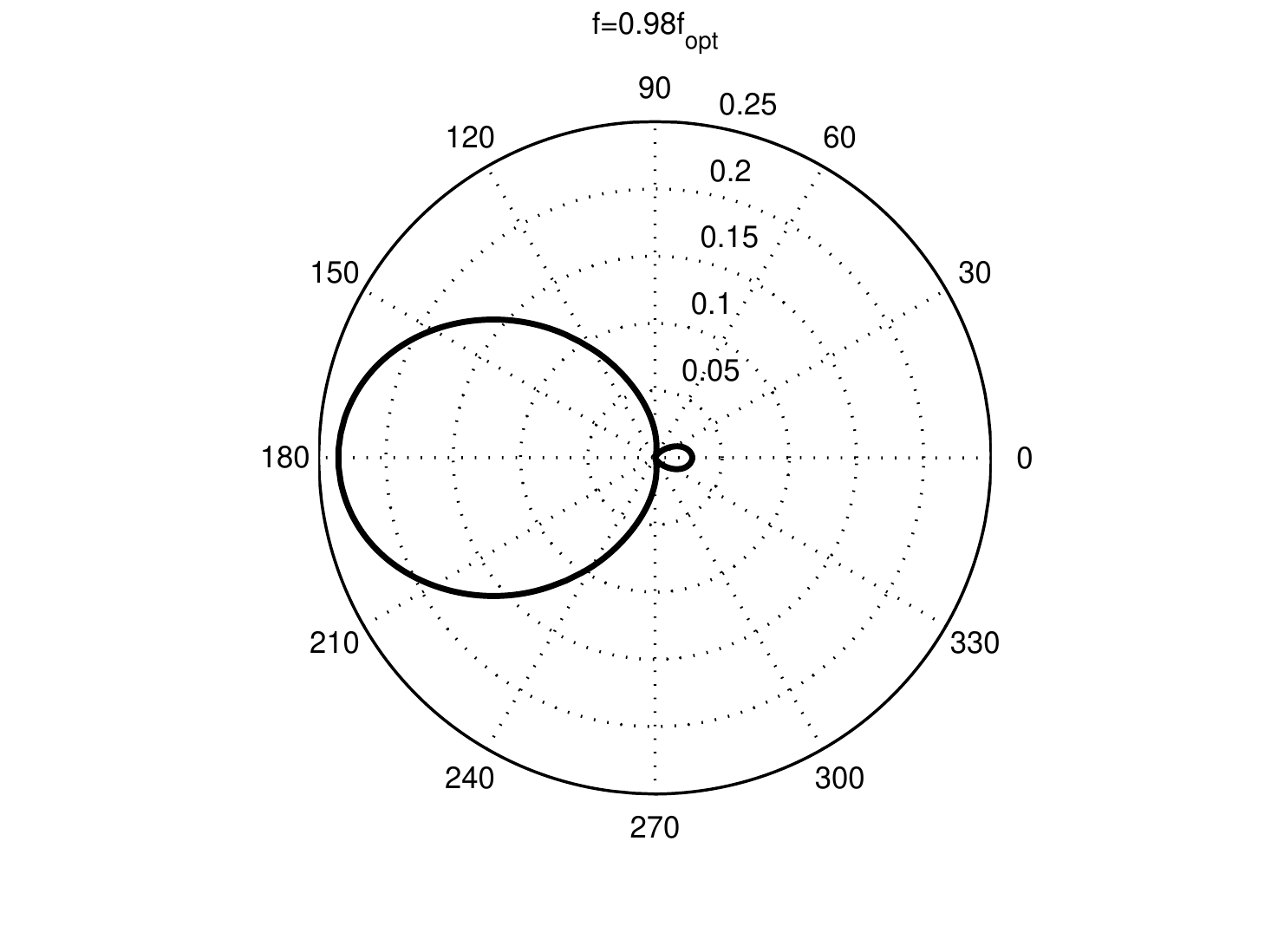}
   \label{fig:Fig3b}}
\subfigure[]{\includegraphics[scale =0.6]{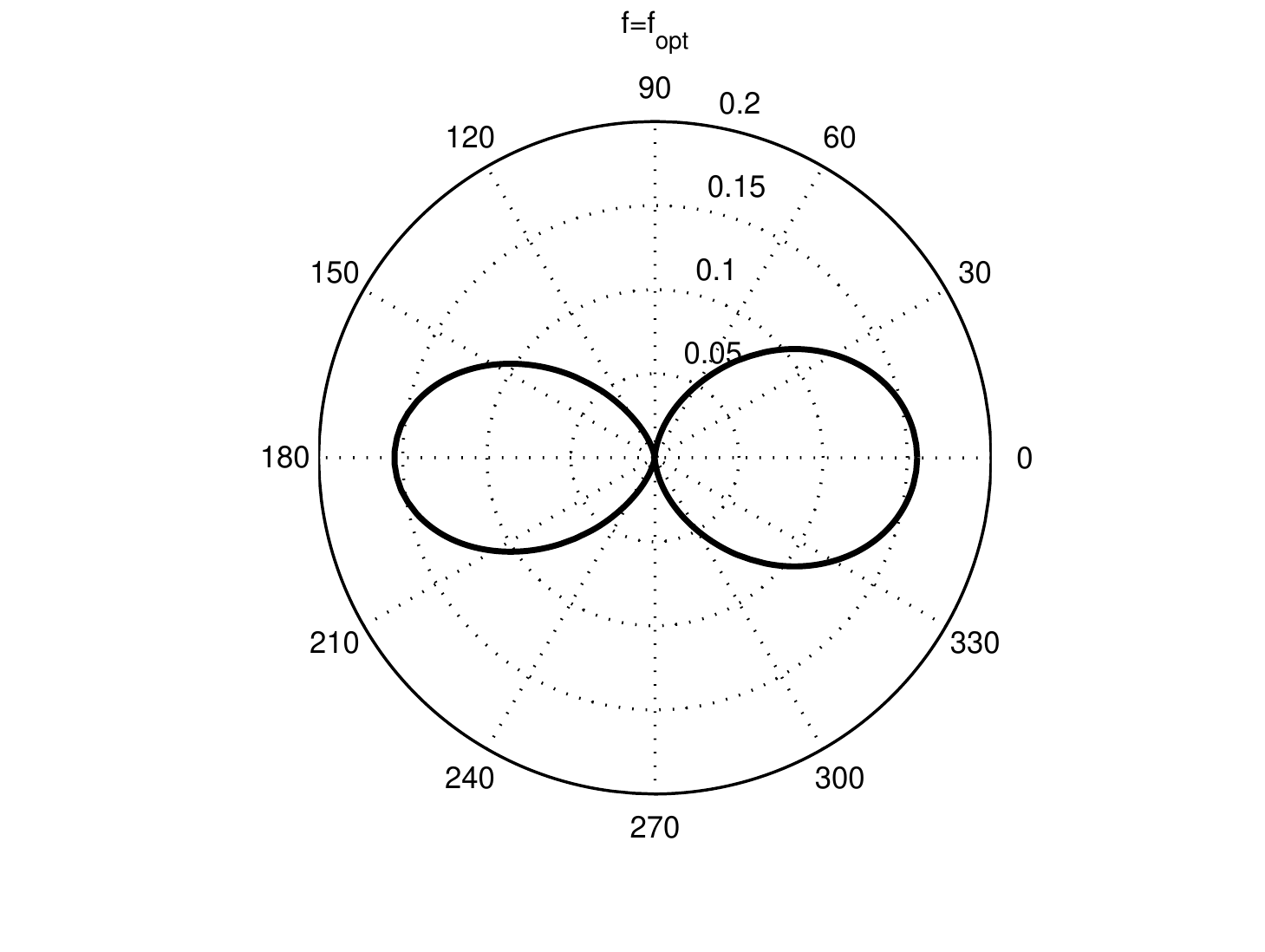}
   \label{fig:Fig3c}}
\subfigure[]{\includegraphics[scale =0.5]{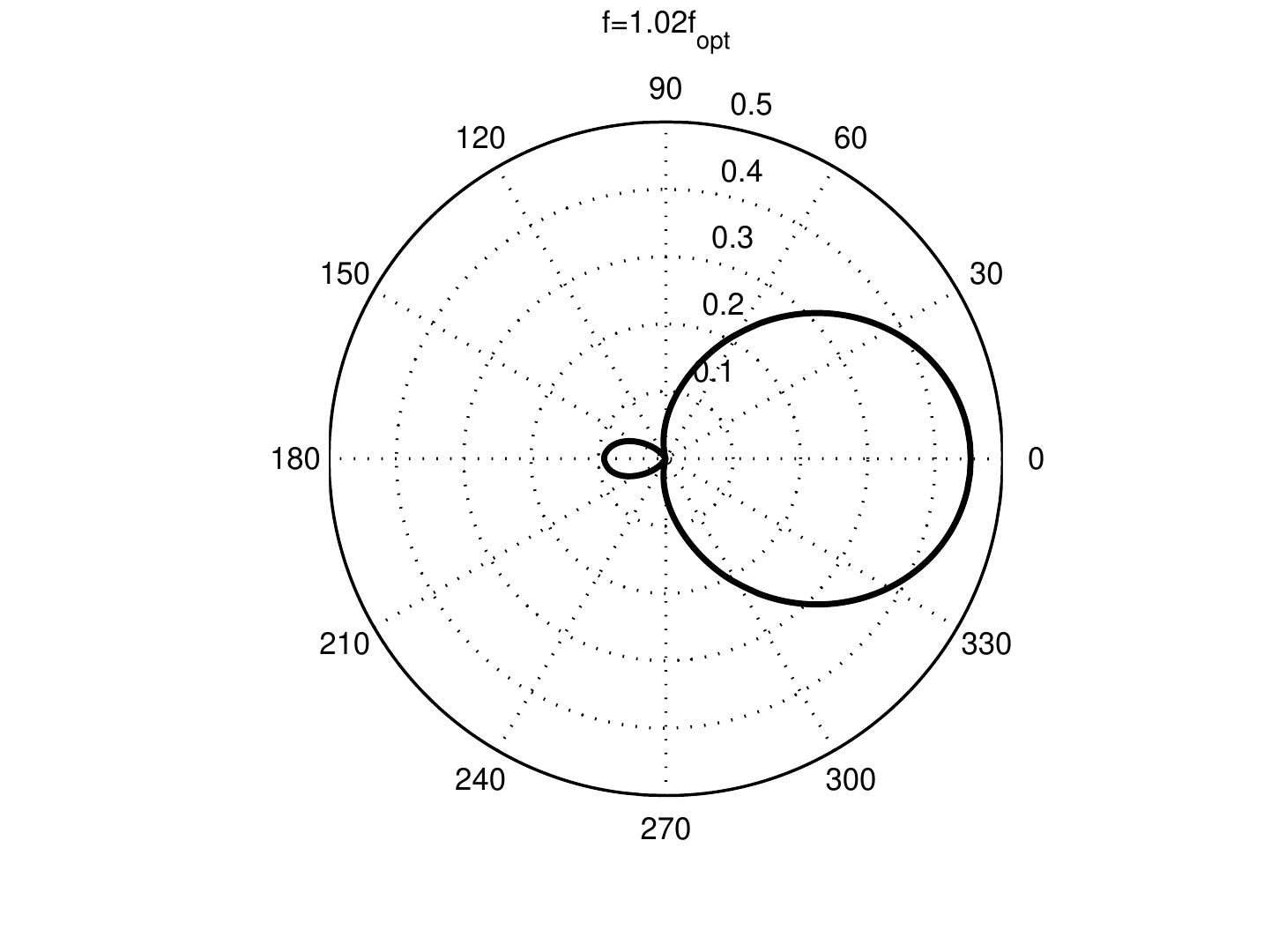}
   \label{fig:Fig3d}}
\subfigure[]{\includegraphics[scale =0.5]{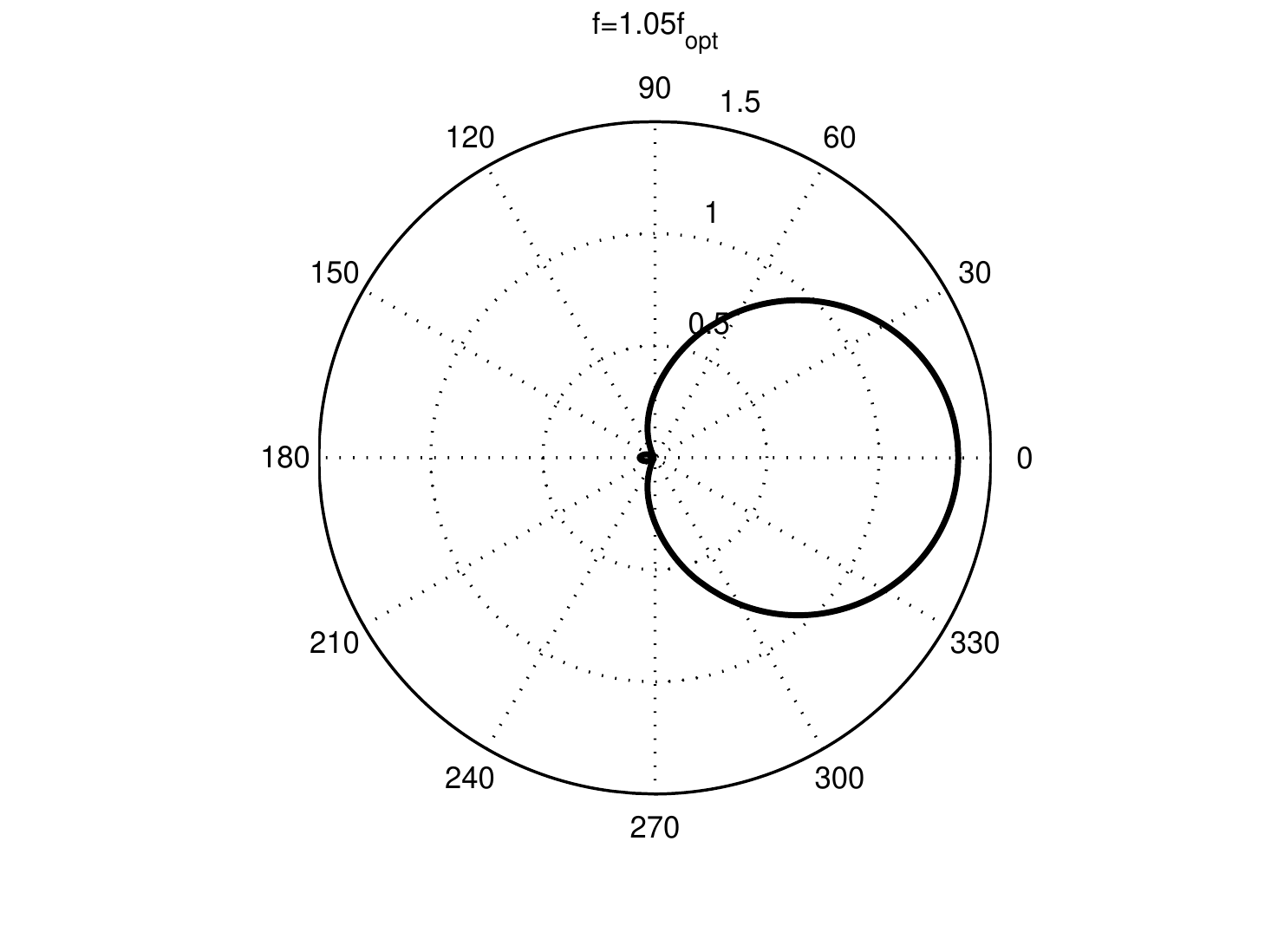}
   \label{fig:Fig3e}}
\caption{The normalized radiation patterns $\lim_{k_0\rho\rightarrow+\infty}\frac{|E_{z0,scat}(\rho, \phi)|}{|\tilde{E}_{z0,scat}(\rho, \phi)|}$ of the cylindrical structure determined from the exact field solution for: (a) $f=0.95f_{opt}$, (b) $f=0.98f_{opt}$, (c) $f=f_{opt}$, (d) $f=1.02f_{opt}$ and (e) $f=1.05f_{opt}$. Plot parameters: $f_0=c/\lambda_0=3\cdot 10^8$ Hz, $f_{opt}\cong 0.992f_0$, $g/\lambda_0=0.05$, $a/\lambda_0=0.08$, $\epsilon_r=60$.}
\label{fig:Figs3}
\end{figure}

In Figs.~\ref{fig:Figs3}, we ``zoom'' in the operation of the device
close to the optimal frequency $f=f_{opt}$ to better understand the
cloaking effect. In particular, we assemble five graphs of the
normalized scattering patterns of the cylindrical structure, defined
by $\lim_{k_0\rho\rightarrow+\infty}\frac{|E_{z0,scat}(\rho,
\phi)|}{|\tilde{E}_{z0,scat}(\rho, \phi)|}$ for $f/f_{opt}=0.95,
0.98, 1, 1.02, 1.05$. We observe that the polar plots differ
substantially from each other despite the smallness of variations of
$f$. More specifically, a backward scattering for $f<f_{opt}$ and a
forward scattering effect for $f>f_{opt}$ has been recorded. Exactly
at $f=f_{opt}$, we obtain a bipolar plot which means that the
first $(n=1)$ term in series (\ref{eq:ElectricField0}), (\ref{eq:ScatteringFarField}) 
becomes dominant in defining the scattering pattern. To
put it alternatively, the omnidirectional zeroth term $(n=0)$ is suppressed
at $f=f_{opt}$ and increases rapidly when going away from this
point. Such behavior reminds us clearly of the scattering
cancellation principle \cite{alu_ScatteringCancellation} which is
reported as being not feasible for PEC cores and the considered TE
(referred to as TM$_z$ in \cite{alu_Retrospection}) polarization.
Consequently, the present cloaking technique can be characterized as
an ``expanded scattering cancellation'' suitable for cloaking of PEC
cylinders. In fact, we have achieved a $de~facto$ scattering cancellation 
effect for PEC objects. Such a feature has been briefly referred in \cite{aluMaterials}.

\begin{figure}[ht]
\centerline{\includegraphics[scale=0.8]{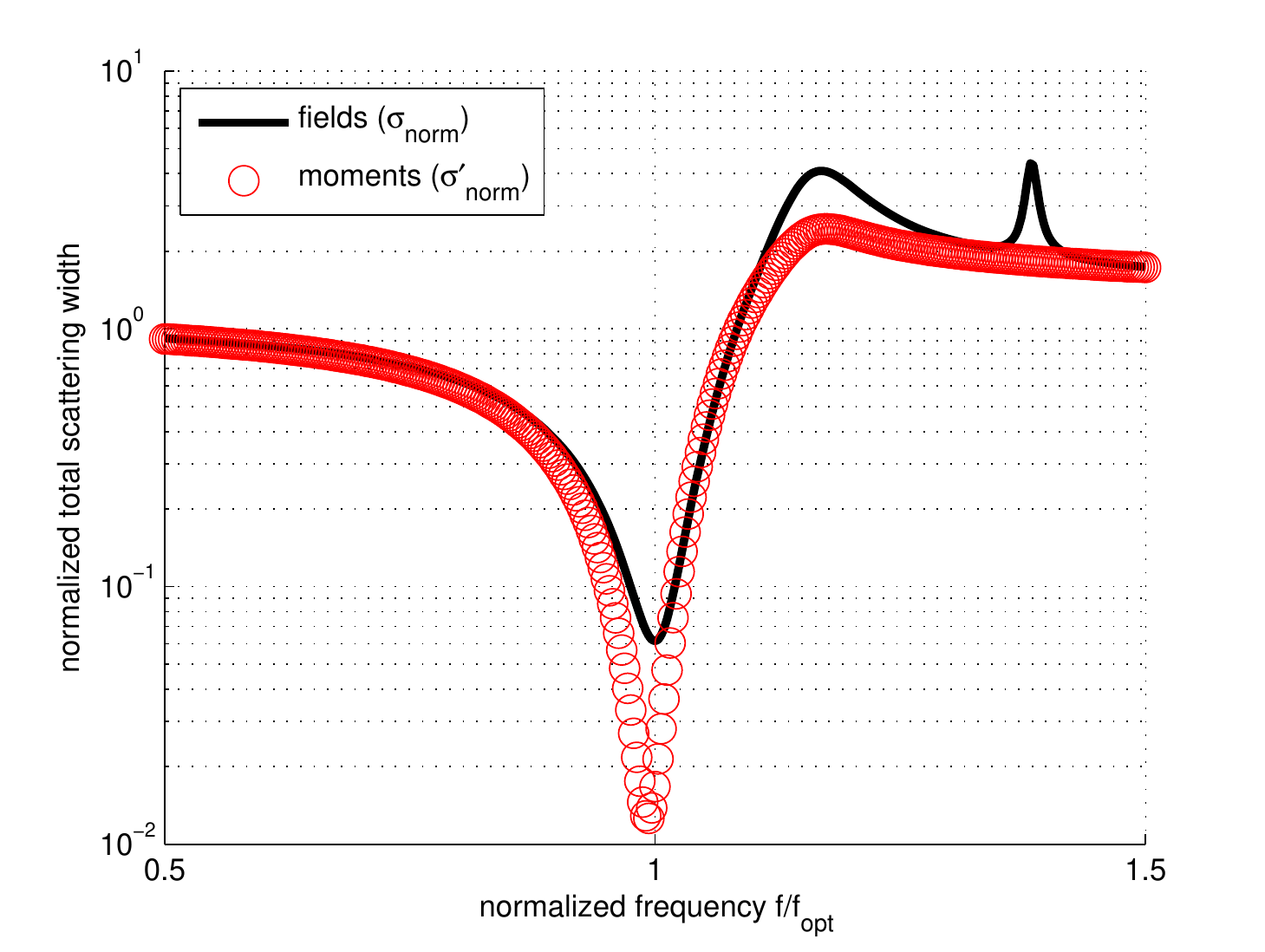}}
\caption{The normalized total scattering width as function of the normalized operating frequency $f/f_{opt}$ when evaluated through the exact field solution $(\sigma_{norm})$ and when computed from the approximate moments model $(\sigma'_{norm})$. Plot parameters: $f_0=c/\lambda_0=3\cdot 10^8$ Hz, $f_{opt}=0.992f_0$, $g/\lambda_0=0.05$, $a/\lambda_0=0.08$, $\epsilon_r=60$.}
\label{fig:Fig4}
\end{figure}

To study the phenomena in the vicinity of the optimal cloaking
frequency $f=f_{opt}$, we adopt the simplified analytical approach with the
equivalent moments described in Section~III. We represent the
variation of the normalized total scattering width when computed
with the exact field formulas ($\sigma_{norm}$) and when evaluated
from the respective moments ($\sigma'_{norm}$) in the same diagram
(Fig.~\ref{fig:Fig4}), as functions of the normalized frequency
$f/f_{opt}$. The former graph is the same as Fig.~\ref{fig:Fig2b},
and the latter exhibits remarkable agreement with the first one,
especially for smaller frequencies. In the vicinity of the cloaking regime, 
the model certainly ``captures'' the physics in terms of the minimized 
scattering; implicitly, optimal frequency $f'_{opt}$ in the moments case is 
only slightly lower than $f_{opt}$ ($f'_{opt}=0.984f_0$). However, the exact 
level of the maximum reduction cannot be predicted successfully since $\sigma'_{norm}$ 
is much smaller than the actual $\sigma_{norm}$. This difference is due to the fact that the higher-order terms $(n>1)$ in (\ref{eq:ElectricField0}), (\ref{eq:ScatteringFarField}) are significant  
at $f\cong f_{opt}, f'_{opt}$ where the scattering due to the leading term is mitigated. Their effect is mutually canceled in the radiation pattern of 
Fig.~\ref{fig:Fig3c} but they indeed substantially contribute to the overall scattered power and define this gap 
between the levels of $\sigma_{norm}$ and $\sigma'_{norm}$ as shown in Fig.~\ref{fig:Fig4}.
Moreover, when the operational frequency gets higher, the moments model becomes inaccurate since it replaces
the entire structure by a pair of dipoles. In other words,
electrically thick cylinders, that imply strong azimuthal variations of induced polarization,
cannot be fitted with the few degrees of freedom of our approximate
moments model. However, the model results exhibit satisfying similarity with the actual field
waveforms at the moderate frequencies and imitate successfully the developed physical mechanism 
at the cloaking regime.

\begin{figure}[ht]
\centering
\subfigure[]{\includegraphics[scale =0.5]{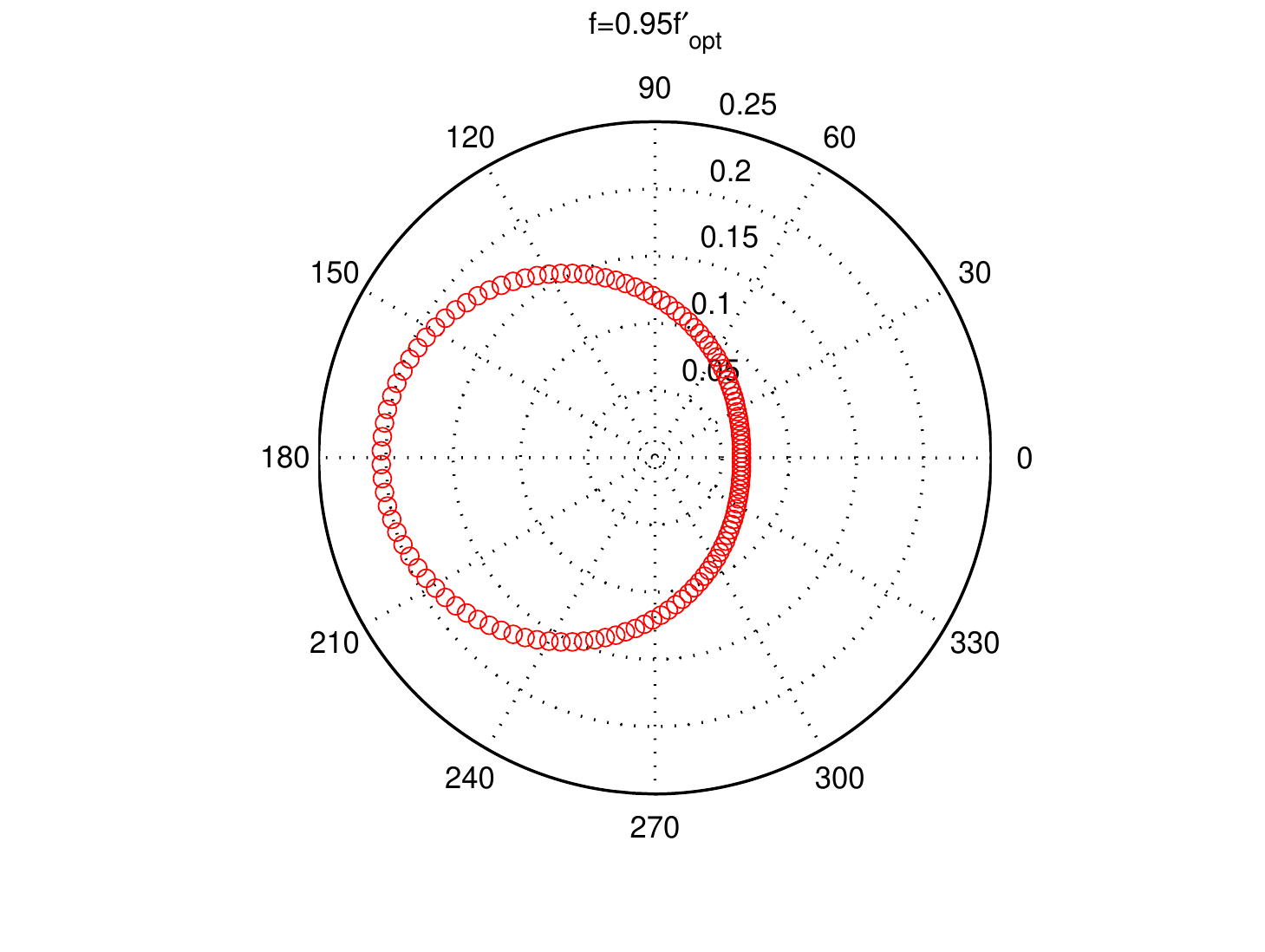}
   \label{fig:Fig5a}}
\subfigure[]{\includegraphics[scale =0.5]{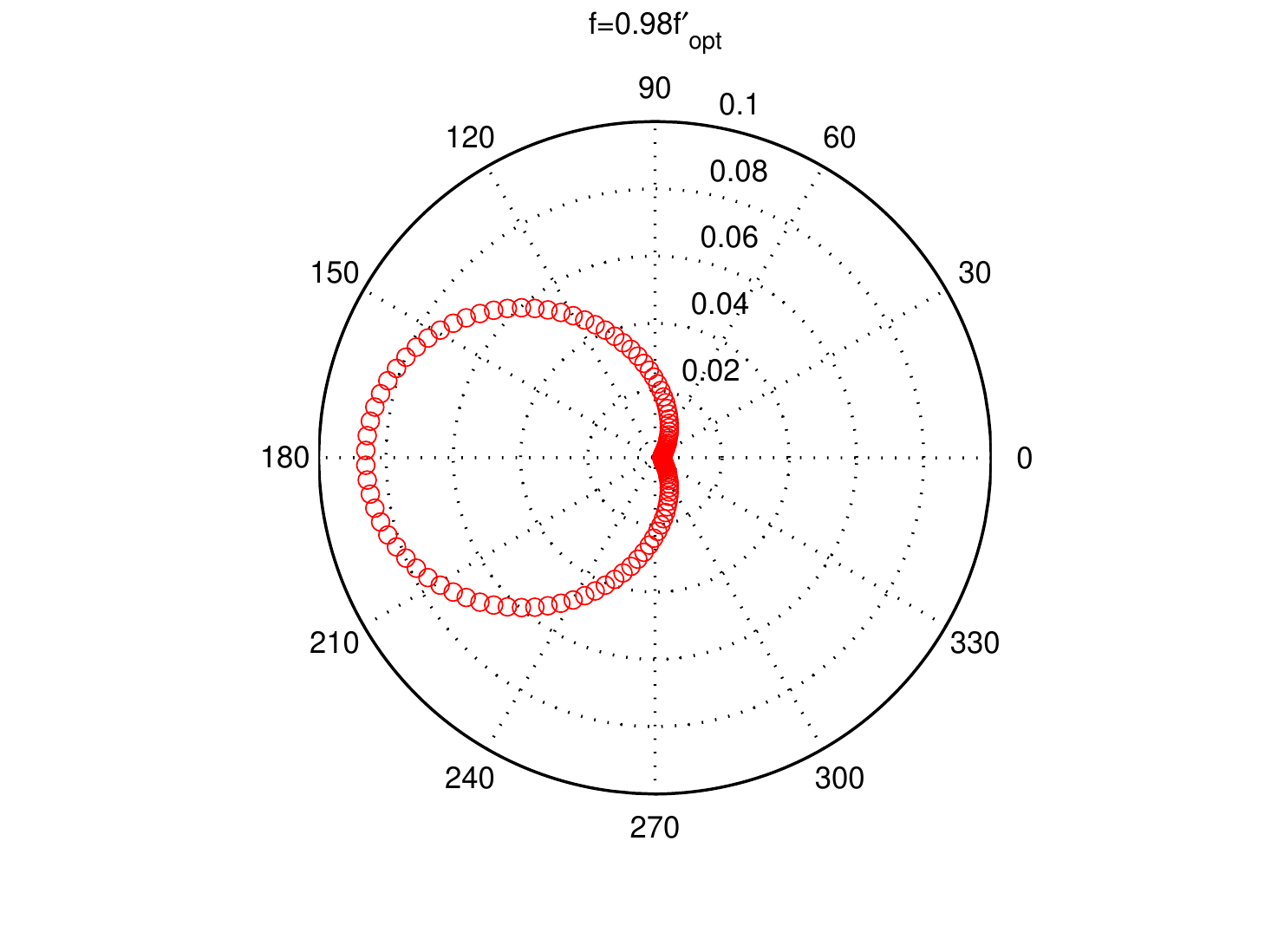}
   \label{fig:Fig5b}}
\subfigure[]{\includegraphics[scale =0.6]{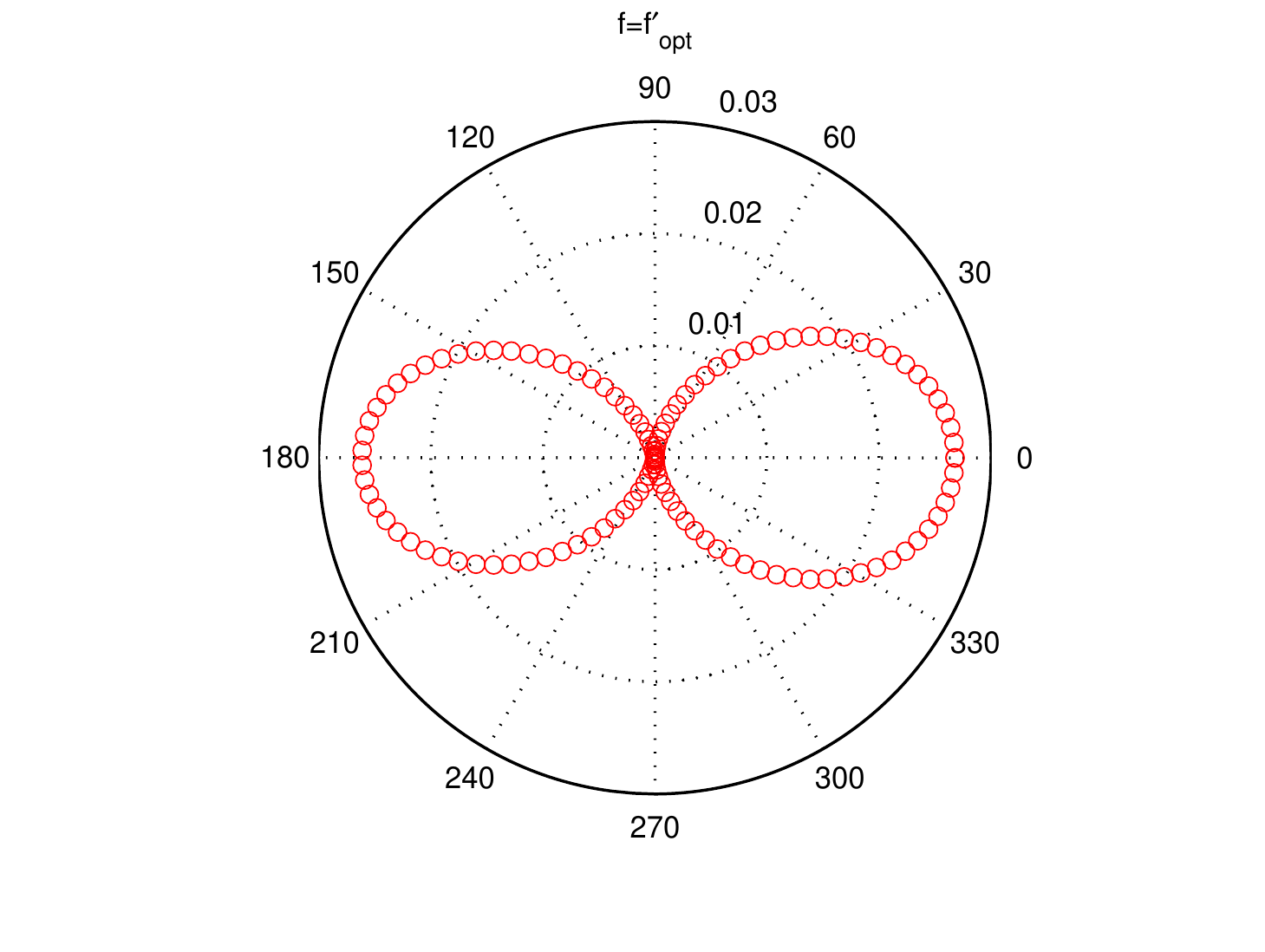}
   \label{fig:Fig5c}}
\subfigure[]{\includegraphics[scale =0.5]{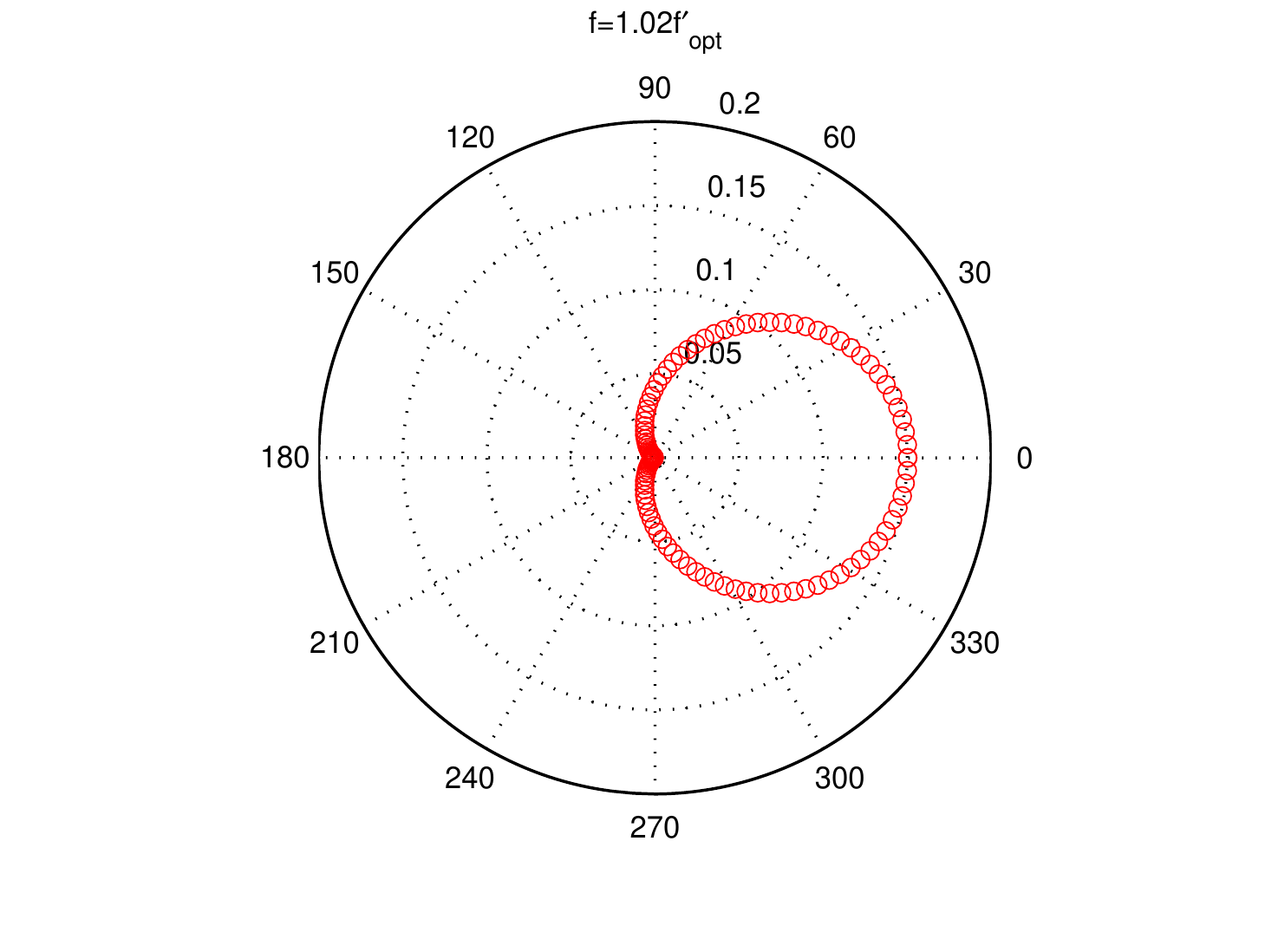}
   \label{fig:Fig5d}}
\subfigure[]{\includegraphics[scale =0.5]{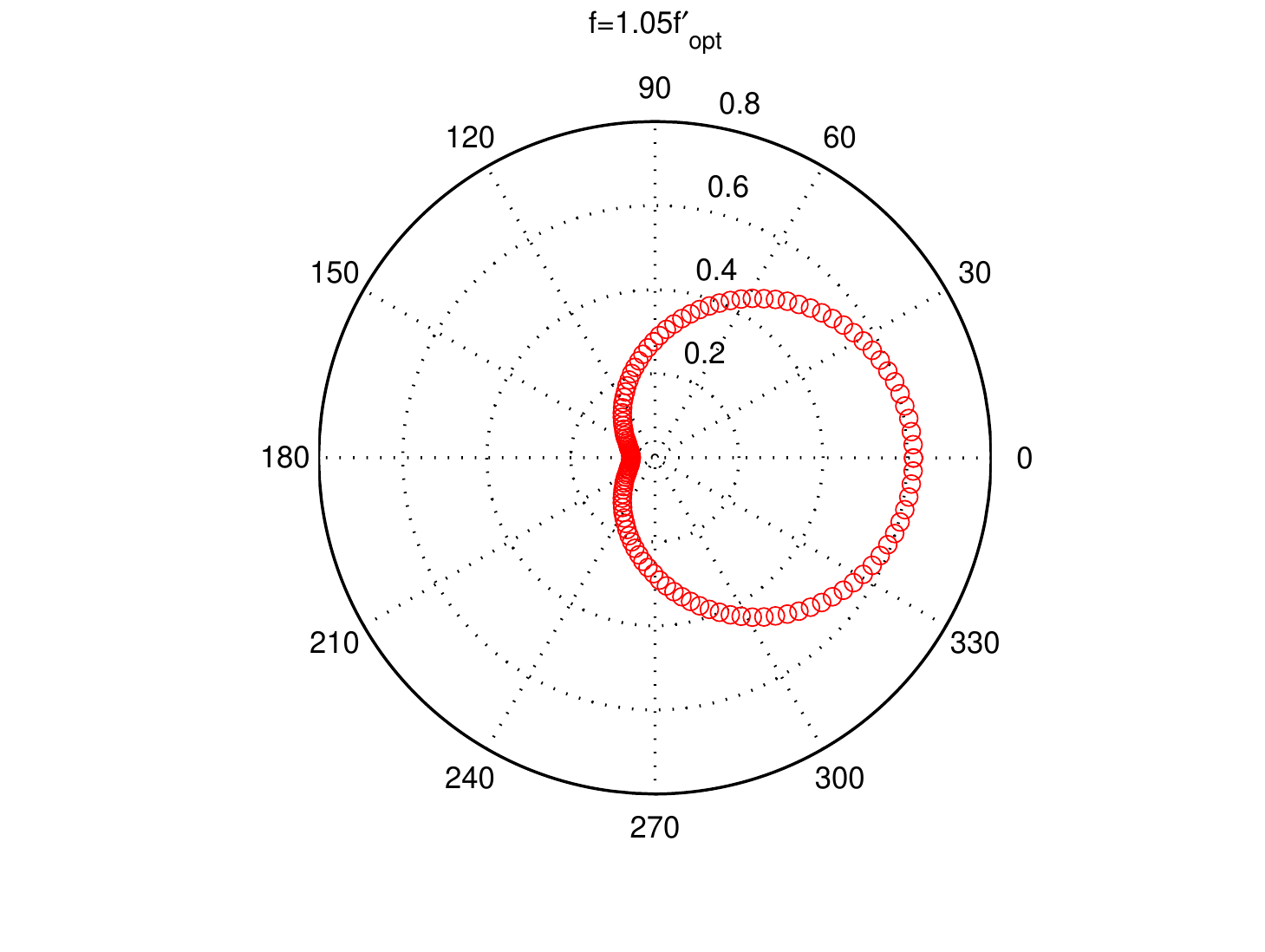}
   \label{fig:Fig5e}}
\caption{The normalized radiation patterns $\lim_{k_0\rho\rightarrow+\infty}\frac{|E'_{z0,scat}(\rho, \phi)|}{|\tilde{E}'_{z0,scat}(\rho, \phi)|}$ of the cylindrical structure determined from the  approximate moments model for: (a) $f=0.95f'_{opt}$, (b) $f=0.98f'_{opt}$, (c) $f=f'_{opt}$, (d) $f=1.02f'_{opt}$ and (e) $f=1.05f'_{opt}$. Plot parameters: $f_0=c/\lambda_0=3\cdot 10^8$ Hz, $f'_{opt}\cong 0.984f_0$, $g/\lambda_0=0.05$, $a/\lambda_0=0.08$, $\epsilon_r=60$.}
\label{fig:Figs5}
\end{figure}

Let us now analyze the Figs.~\ref{fig:Figs5}, which are analogous to 
Figs.~\ref{fig:Figs3} but with the use of the moments model 
(the quantity represented is defined as: $\lim_{k_0\rho\rightarrow+\infty}\frac{|E'_{z0,scat}(\rho,
\phi)|}{|\tilde{E}'_{z0,scat}(\rho, \phi)|}$). Again, the cloaking
effect is observed during the transition from the backward to
forward scattering in the radiation pattern of the structure.
Exactly on the optimal frequency $f'_{opt}$ the coated cylinder
radiates as a magnetic dipole line, namely only the term
proportional to $m_y^*$ in (\ref{eq:RadiatedFarField}) is of
significant magnitude. In this way, we reach the conclusion that the other
term of (\ref{eq:RadiatedFarField}) corresponding to the electric moment 
$c p_z^*$ is suppressed. In this way, it becomes evident that our model combines 
the electric and magnetic effect into one formula and assigns to each of them a 
different azimuthal profile: omnidirectional and bipolar respectively.  
As reported in Fig. \ref{fig:Fig4}, the values of the scattering response are
much lower than in the exact case.

\begin{figure}[ht]
\centerline{\includegraphics[scale=0.8]{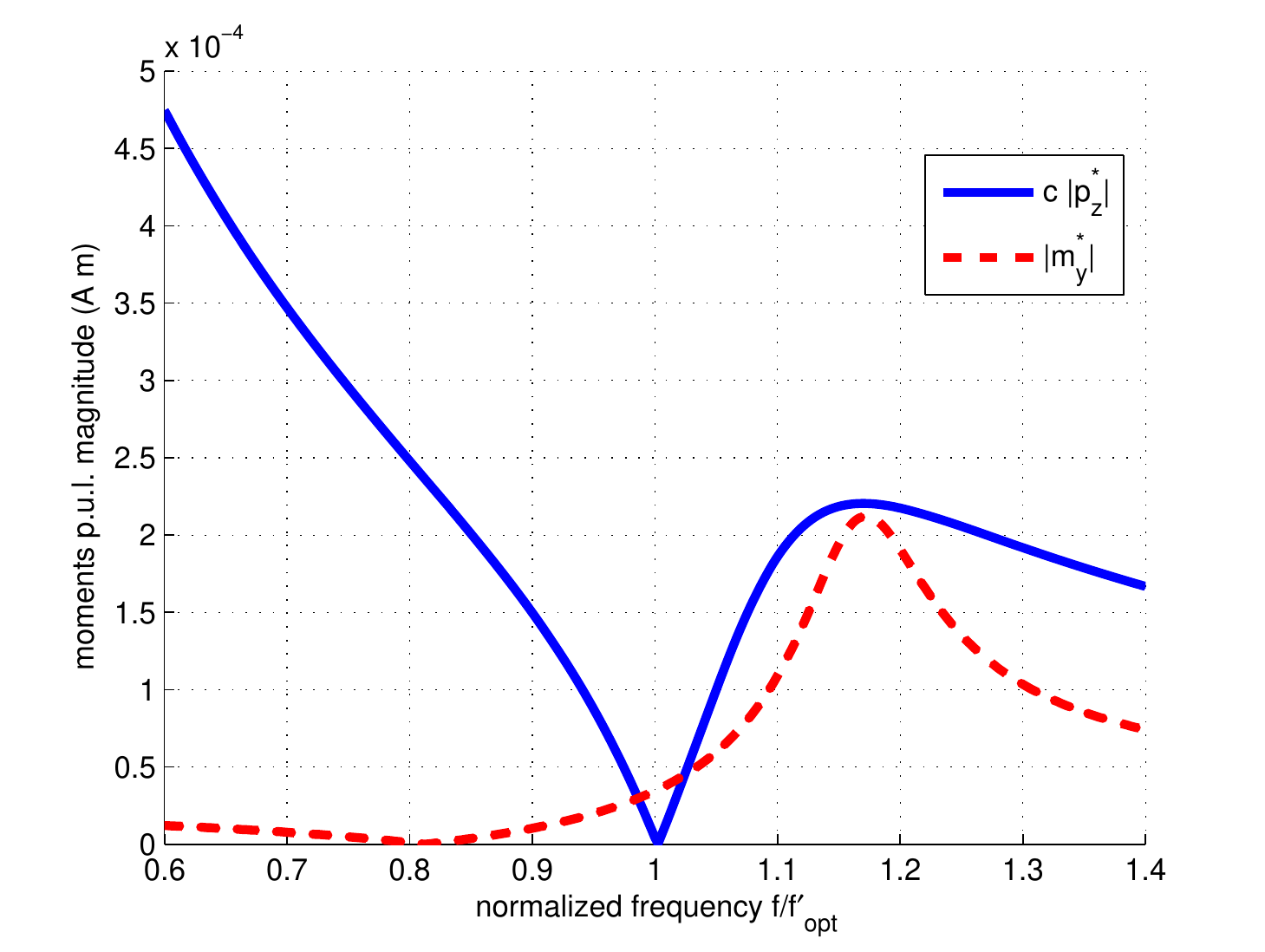}}
\caption{The magnitudes of the moments per unit length (p.u.l.) of axis $\textbf{z}$ ($c~p^*_z, m^*_y$) as functions of the normalized operating frequency $f/f'_{opt}$. Plot parameters: $f_0=c/\lambda_0=3\cdot 10^8$ Hz, $f'_{opt}=0.984f_0$, $g/\lambda_0=0.05$, $a/\lambda_0=0.08$, $\epsilon_r=60$.}
\label{fig:Fig6}
\end{figure}

After validation of the model in the vicinity of the cloaking frequency $f'_{opt}\cong f_{opt}$,
we can make exclusive use of it hereinafter. 
In Fig.~\ref{fig:Fig6} we show the magnitude variation
of the two amplitudes incorporated in (\ref{eq:RadiatedFarField}).
The first one $(c|p_z^*|)$, corresponding to the electric response
of the structure, is minimized (almost nullified) at $f=f'_{opt}$ as 
predicted above. The second one $(|m_y^*|)$, determining the magnetic response of the
cylinder, increases smoothly in the neighborhood of $f'_{opt}$. 
Again, it is clearly noted that the electric moment represents the 
omnidirectional term of the far-field
pattern, while the magnetic moment is proportional to the bipolar
term as defined by expression (\ref{eq:RadiatedFarField}).
Therefore, the findings from the graphs are in accordance with
those of Figs.~\ref{fig:Figs3} (based on the explicit field
formulas), since the omnidirectional component is mitigated at the
cloaking frequency.

\begin{figure}[ht]
\centering
\subfigure[]{\includegraphics[scale =0.5]{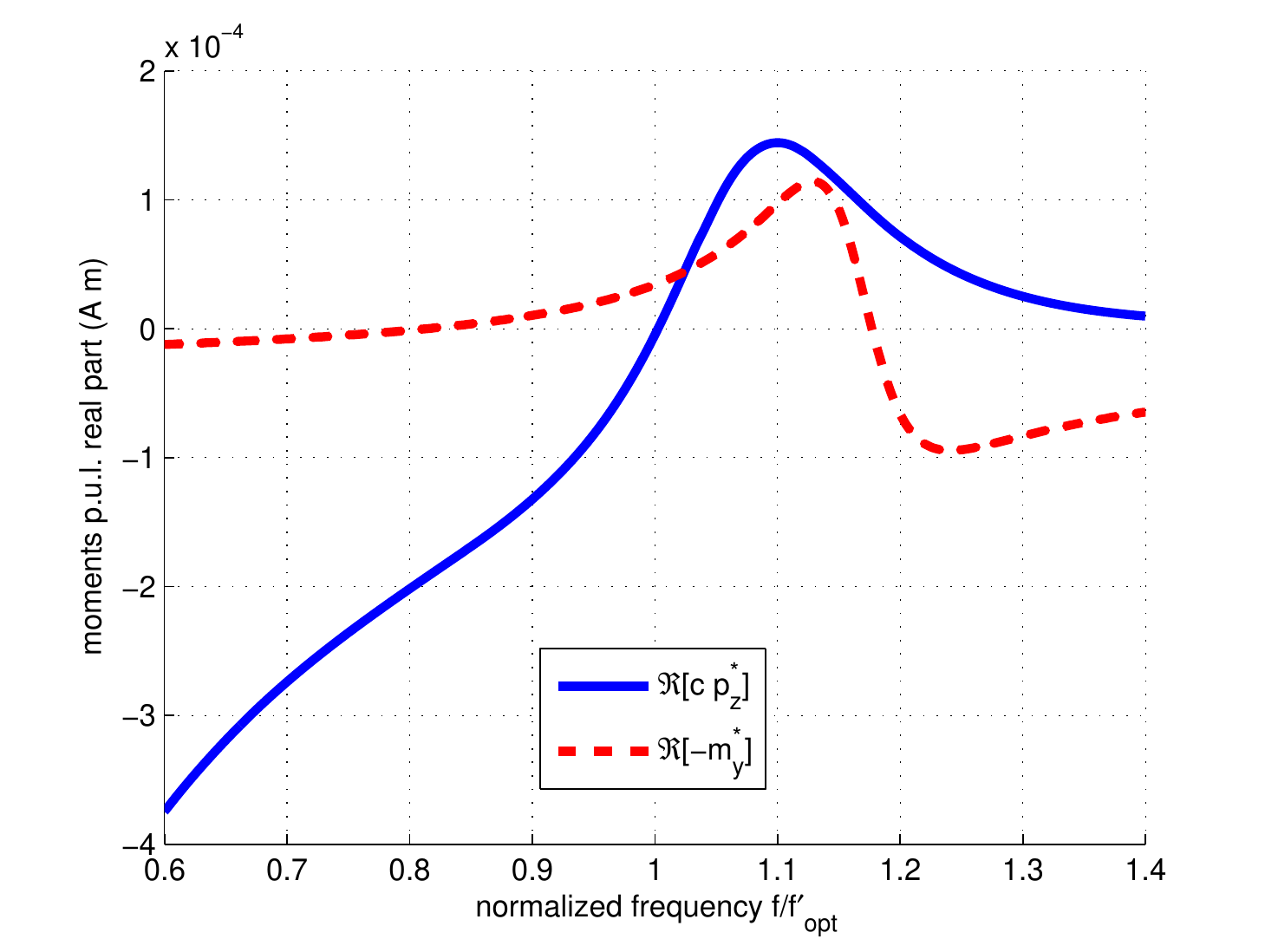}
   \label{fig:Fig7a}}
\subfigure[]{\includegraphics[scale =0.5]{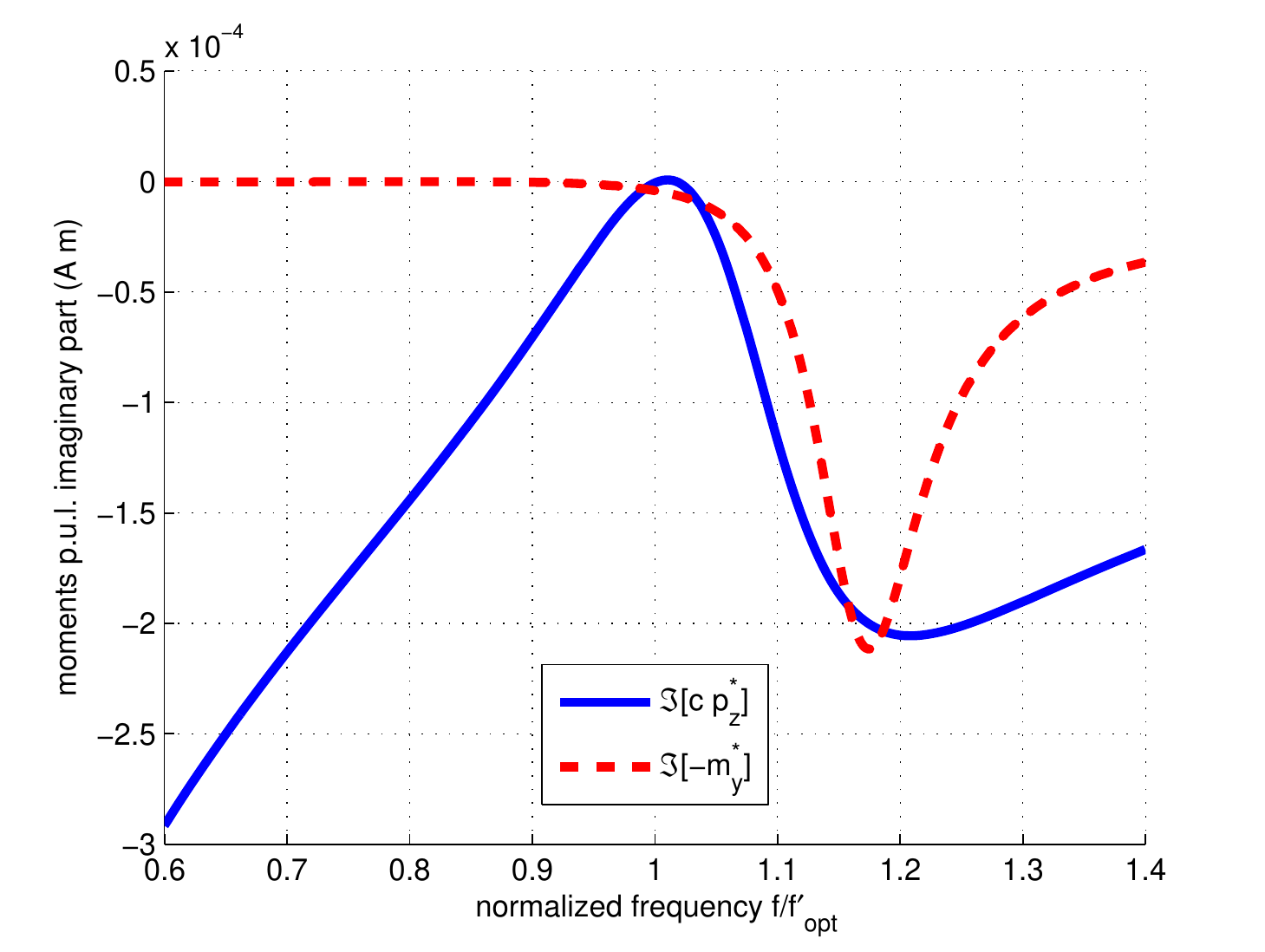}
   \label{fig:Fig7b}}
\caption{The: (a) real and (b) the imaginary parts of the moments $c~p^*_z$ and $(-m^*_y)$ as functions of the normalized operating frequency $f/f'_{opt}$. Plot parameters: $f_0=c/\lambda_0=3\cdot 10^8$ Hz, $f'_{opt}=0.984f_0$, $g/\lambda_0=0.05$, $a/\lambda_0=0.08$.}
\label{fig:Figs7}
\end{figure}

In Figs.~\ref{fig:Figs7}, we show the variations of the constituent parts of the quantities $c p_z^*$ and $(-m_y^*)$ 
(real in Fig. \ref{fig:Fig7a} and imaginary in Fig. \ref{fig:Fig7b}) with respect to 
the normalized operational frequency $f/f'_{opt}$. The negative sign in the magnetic moment is compatible with (\ref{eq:RadiatedPower}), where
the scattering power is written as the result of the action of two sources: one with magnitude $\Im[c p_z^*]$ and the other with magnitude $\Im[-m_y^*]$. From Fig. \ref{fig:Fig7b}, it is apparent that the two quantities cannot give a zero sum at any frequency, since the imaginary parts (responsible for scattering losses) retain the same sign (negative) at all frequencies. Therefore, it is sensible for the optimal choice to be $f\cong f'_{opt}$ since: (i) both imaginary parts are small as demonstrated in Fig. \ref{fig:Fig7b}, (ii) the real part $\Re[c p_z^*]$ is nullified at $f=f'_{opt}$ and its magnitude increases rapidly slightly far from it as shown in Fig. \ref{fig:Fig7a} and (iii) the real part $\Re[-m_y^*]$ does not vary considerably in the vicinity of $f=f'_{opt}$ as again indicated in Fig. \ref{fig:Fig7a}. When observing the frequency dispersion of the real and the imaginary parts of the electric moment, one concludes that it exhibits a typical plasma dependence tending to large negative values at low frequencies. On the other hand, the dispersion of the magnetic moment resembles a Lorentz-type curve with a resonance at higher frequencies.

\begin{figure}[ht]
\centering
\subfigure[]{\includegraphics[scale =0.5]{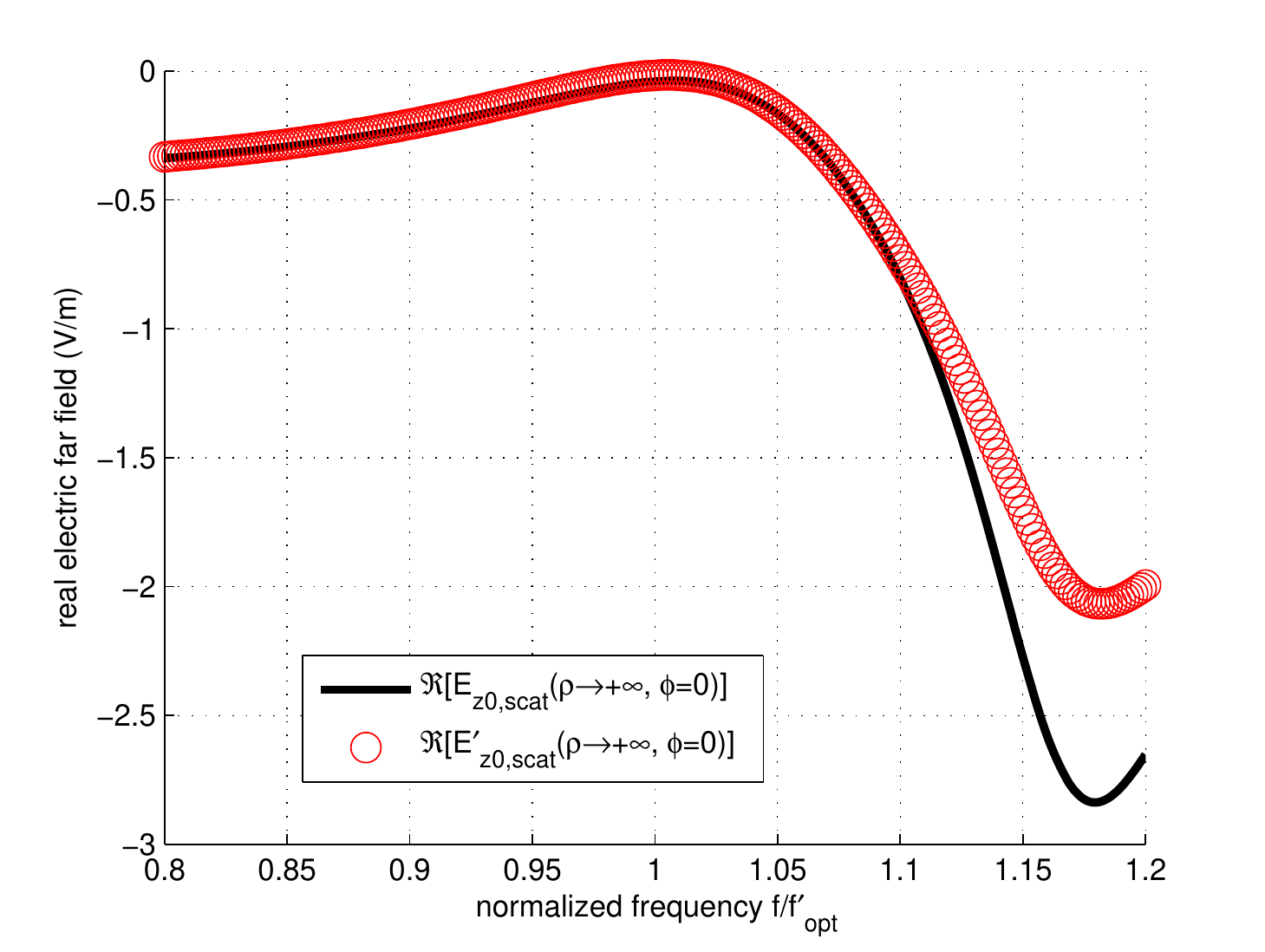}
   \label{fig:Fig8a}}
\subfigure[]{\includegraphics[scale =0.5]{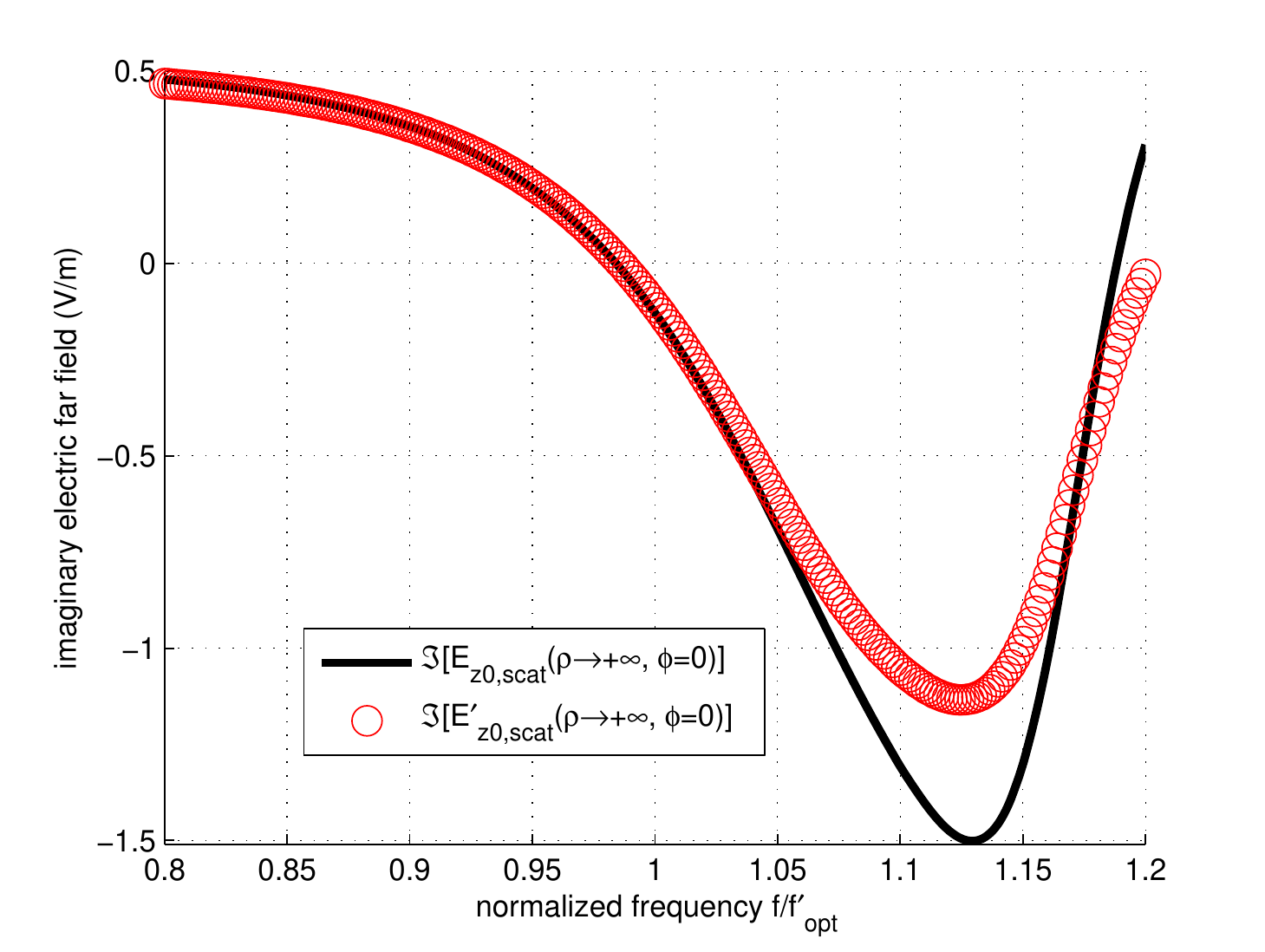}
   \label{fig:Fig8b}}
\caption{The: (a) real and (b) the imaginary parts of the scattered electric field in the far region along the forward direction $(\phi=0)$, using exact formulas ($\lim_{k_0\rho\rightarrow+\infty}E_{z0,scat}(\rho, \phi=0)$) and the approximate moments model ($\lim_{k_0\rho\rightarrow+\infty}E'_{z0,scat}(\rho, \phi=0)$), as functions of the normalized operating frequency $f/f'_{opt}$. Plot parameters: $f_0=c/\lambda_0=3\cdot 10^8$ Hz, $f'_{opt}=0.984f_0$, $g/\lambda_0=0.05$, $a/\lambda_0=0.08$.}
\label{fig:Figs8}
\end{figure}

In Figs.~\ref{fig:Figs8}, we represent the real and the imaginary parts of the scattering field in the forward direction ($\phi=0$) in the far region as functions of $f/f'_{opt}$ from expressions (\ref{eq:ScatteringFarField}), (\ref{eq:RadiatedFarField}), where the common factor $\sqrt{\frac{2}{\pi k_0\rho}}e^{-j(k_0\rho-\frac{\pi}{4})}$ has been dropped. By taking into account that the scattered power by the entire cylindrical structure is proportional to the real part of the electric far field as stated in (\ref{eq:ScatteredPower}) and observing Fig. \ref{fig:Fig8a}, one concludes again that the perfect cloaking is not possible and only a minimization of the overall response is feasible. As far as the imaginary part of the forward-scattering quantity is concerned, it is kept moderate for $f\cong f'_{opt}$ regardless of the method (exact or approximate) one employs to evaluate it.

\section{Conclusions}
An analytical model based on the equivalent moments of a cylindrical PEC rod covered by a conventional dielectric is used to analyze the minimal scattering response of the structure when it is illuminated by a normally incident plane wave (with electric field parallel to the cylinder's axis). The adopted approximate model is reliable for moderate electrical sizes and captures the scattering effect of the illuminated object. In the vicinity of the cloaking frequency, a drastic change in the radiation pattern of the covered PEC cylinder is observed; in particular, there is a clear transition from  backward to forward scattering. It has been found that the minimum scattering is achieved when the electric moment is mitigated and in that case the system radiates as a magnetic dipole line.

\newpage

\end{document}